\documentclass[twocolumn,twocolappendix]{emulateapj}

\usepackage{graphicx,times}
\usepackage{longtable}
\usepackage{array}
\usepackage[colorlinks,dvipdfm,linkcolor=blue,anchorcolor=,citecolor=blue]{hyperref}

\shorttitle{Numerical integral of resistance coefficients in diffusion}
\shortauthors{Q. S. Zhang}

\begin{document}

\title{Numerical integral of resistance coefficients in diffusion}
\author{Q. S. Zhang \altaffilmark{1,2}}
\email{zqs@ynao.ac.cn(QSZ)}

\altaffiltext{1}{Yunnan Observatories, Chinese Academy of Sciences, P. O. Box 110, Kunming 650011, China.}
\altaffiltext{2}{Key Laboratory for the Structure and Evolution of Celestial Objects, Chinese Academy of Sciences, Kunming, 650011, China.}

\begin{abstract}

The resistance coefficients in screen Coulomb potential of stellar plasma are evaluated in high accuracy. I have analyzed the possible singularities in the integral of scattering angle. There are possible singularities in the case of attractive potential. This may result in problem for numerical integral. In order to avoid the problem, I have used a proper scheme, e.g., splitting into many subintervals and the width of each subinterval is determined by the variation of the integrand, to calculate the scattering angle. The collision integrals are calculated by using Romberg's method therefore the accuracy is high (i.e., $ \sim 10^{-12}$). The results of collision integrals and their derivatives in $-12 \leq \psi \leq 5$ are listed. By using Hermite polynomial interpolation from those data, the collision integrals can be obtained with an accuracy of $10^{-10}$. For very weak coupled plasma ($\psi \geq 4.5$), analytical fittings for collision integrals are available with an accuracy of $10^{-11}$. I have compared the final results of resistance coefficients with other works and found that, for repulsive potential, the results are basically same to others, for attractive potential, the results in intermediate and strong coupled case show significant differences. The resulting resistance coefficients are tested in the solar model. Comparing with the widely used Cox et al.(1989) and Thoul et al. (1994) models, the resistance coefficients in screen Coulomb potential leads to a little weaker effect in solar model, which is contrary to the expectation of attempts to solve the solar abundance problem.

\end{abstract}


\section{Introduction}

On the transport of elements in stellar interior, beside the macro processes (e.g., convective mixing and rotation induced mixing), there are micro transport processes induced by gradients of pressure, temperature and elements abundances, and the radiation pressure, which are called micro diffusion in collective. The pressure gradient (equivalent to gravity) and the temperature gradient tend to concentrate heavier elements toward the stellar center. The elements abundances gradients tend to form a chemical homogeneous structure, which is contrary to the gradients of pressure and temperature. The effect of radiation pressure on ions are called the radiation acceleration, which leads to a radiative acceleration determined by the state of ionization of concerned species \citep{mi91}. The movements of particles of a species are determined by the values of the radiative acceleration the gravitational acceleration. For completely ionized species, the radiative accelerations are negligible. In this paper, the radiation acceleration is not taken into account.

General treatments for the micro diffusion in a multi-component fluid have been provided by Burgers equations \citep{bur69} and Chapman-Enskog theory \citep{cc70}. Those two theories used different approximations in handling the Boltzmann equation, but they are ultimately equivalent to each other in the limit of collision-dominated plasma \citep{bur69}. The diffusion velocity of each species can be obtained by solving Burgers equations. In Burgers equations, there are resistance coefficients $K_{ij}$, $z_{ij}$, $z_{ij}'$ and $z_{ij}''$ which represent the effect of collisions between charged particles. Their precise evaluation is required for correct estimate of the detail of diffusion process of different elements in stellar interior. \citet{bur69} have calculated the resistance coefficients in a pure Coulomb potential with a long-range cut-off. In stellar interior, a better description of the collisions between charged particles is to used the Debye-H\"{u}ckel type of potential which takes into account the effect of screening. The resistance coefficients are determined by collision integrals. Collision integrals in screening potential have been studied in many works \citep[e.g.,][]{mms67,m84,im85,ppfm86}. In the calculations of \citet{mms67}, the range is smaller than that of \citet{ppfm86}. \citet{m84} has only calculated the case of repulsive potential and the results in the case of weak coupled plasma seems numerical instable. \citet{ppfm86} have calculated the collision integrals in large range of the couple parameter $\psi$ and provided fitting formula for the collision integrals. However, their results of collision integrals in the case of attractive potential show significant oscillatory behaviors in the intermediate and strong coupled case. As pointed out by \citet{mac91} and \citet{bf14}, this oscillatory behaviors are caused by inappropriate treatment of an internal singularity in the integral of collision cross section. Another issue in \citet{ppfm86} is that a low order Gauss-Legendre formula has been used in the calculations of scattering angle. This scheme may fail to evaluate the scattering angle in the cases that the integrand of the scattering angle integral has singularity. Recently, \citet{bf14} have used supercomputing and have evaluated the collision integrals for the attractive potential case in an accuracy of $10^{-6}$. 

Although the Debye-H\"{u}ckel type of potential is more reasonable than the truncated Coulomb potential in stellar interior, it is inaccurate in the strongly coupled regime \citet{ppfm86}. Recently, \citet{bd13,bd14} have used the concept of effective interaction potential and calculated it by using the hypernetted-chain approximation. Benchmarked against molecular dynamic simulations, the results based on the hypernetted-chain approximation are widely valid even for very strongly coupled plasmas. The effective potential theory has been used in studying strongly coupled regime \citep{bd15,db16}. In the weakly coupled case, the hypernetted-chain approximation results in the screened Coulomb potential. The differences of collision integrals between the hypernetted-chain approximation and the screened Coulomb potential becomes significant in the strongly coupled regime $\Gamma > 10$ \citep[see e.g.,][Fig.7]{bd14}. For the weakly and intermediate coupled cases, the screened Coulomb potential is accurate enough. 

The micro diffusion is a basic physical process concerned in modern solar evolutionary models, since the differences of the structure of the solar interior between models and helioseismic inversions are significantly reduced by taking into account the micro diffusion \citep{bah95,chr96}. The solar standard models with the old \citet[][GN93]{GN93} and \citet[][GS98]{GS98} compositions are in good agreement with the helioseismic inversions on sound speed, density, surface helium abundance $Y_s$, the base of the solar convection zone $R_{bc}$. However, in the last decade, the updated compositions \citep[e.g.,][]{AGSS09} with low surface metallicity make solar models worse. As the metallicity decreasing, $R_{bc}$ in the solar models becomes larger, $Y_s$ becomes lower, the differences of sound speed and density in the solar interior between models and helioseismic inversion becomes larger \citep[see e.g.,][]{ba04,mon04,bah05,yb07,AGSS09}. This is the so called 'solar abundance problem'. A proposal to this problem is to enhance the micro diffusion in solar interior \citep[e.g.,][]{mon04,gwc05,yb07,y16}, which increases metallicity in the interior of solar models and attempts to make the interior structure of solar model close to the model with old composition GN93 or GS98. A widely used model of micro diffusion in solar models has been developed by \citet[][CGK89]{cgk89} and \citet[][TBL94]{tbl94}. It is found that multiplying a factor $\sim 2$ on the diffusion velocities derived by using CGK89 and TBL94 model can obtain a correct $R_{bc}$ \citep{mon04,gwc05,yb07,y16}. In CGK89 and TBL94 model, the resistance coefficients $z_{ij}$, $z_{ij}'$ and $z_{ij}'$ are based on truncated (long-range cut-off) pure Coulomb potential, but the resistance coefficient $K_{ij}$ is based on a fitting on the numerical results of the repulsive case of the screening Coulomb potential. Therefore the model could be improved by using self-consistent resistance coefficients based on screening Coulomb potential. It is helpful for the solar abundance problem to investigate the effect of micro diffusion with resistance coefficients based on screening Coulomb potential.

The main improvements in this paper are to properly handle singularities in scattering angle integral and to use accurate schemes in the calculations of numerical integrals. Comparing with \citet{ppfm86,mac91}, the accuracy of the collision integrals for both repulsive and attractive cases are improved to $10^{-12}$. In this paper, I have calculated the numerical collision integrals in screening Coulomb potential in high accuracy, compared the resulting resistance coefficients with other works and tested them in the solar model. The formulas of resistance coefficients are introduced in Section \ref{RCformula}, the details of the numerical calculation of collision integrals are presented in Section \ref{numintegral}, the results of collision integrals are shown in Section \ref{results}, the comparison of resulting resistance coefficients with other works are discussed in Section \ref{linktoothers}, the resistance coefficients are tested in the solar model in Section \ref{app}, and Section \ref{summary} is a summary.

\section{Formulas of resistance coefficients in diffusion} \label{RCformula}

The resistance coefficients in Burgers equation are determined by collision integrals. The expression of collision integrals between species $i$ and $j$ is as follow \citep{m84,ppfm86}:
\begin{eqnarray}
&&{\Omega _{ij}}^{(ml)} = \sqrt {\frac{{kT (m_i+m_j) }}{{2\pi {m_i m_j}}}} \int\limits_0^{ + \infty }  {{{\textrm{exp}}{( - {g^2})}}{g^{2l + 3}}{\phi _{ij}}^{(m)}dg} ,
\end{eqnarray}%
with the collision cross sections for given energy
\begin{eqnarray}
&&{\phi _{ij}}^{(m)} = 2\pi \int\limits_0^{ + \infty } {(1 - {{\cos }^m}{\chi _{ij}})bdb} ,
\end{eqnarray}%
and the scattering angle
\begin{eqnarray}
&&{\chi _{ij}} = \pi  - 2\int\limits_{{r_{0,ij}}}^{ + \infty } {\frac{{bdr}}{ {r^2}\sqrt {1 - (b/r)^2 - V_{ij}(r)/ (g^2kT)} }} ,
\end{eqnarray}%
where $m_i$ is the mass of a particle of the $i$-th species and $r_{0,ij}$ is the distance of closest approach for given impact parameter $b$ determined by this equation:
\begin{eqnarray}
&&1 - {(\frac{b}{{{r_{0,ij}}}})^2} - \frac{{{V_{ij}}({r_0})}}{{{g^2}kT}} = 0.
\end{eqnarray}%
In the above equations, $k$ is Boltzmann constant, $T$ is temperature, $g$ is a dimensionless velocity and $V_{ij}(r)$ is the potential.

In the stellar interior, the static screened Coulomb potential is more reasonable than the pure Coulomb potential. The static screened Coulomb potential is given by
\begin{eqnarray}
&&{V_{ij}}(r) = {Z_i}{Z_j}{e^2}\frac{{\exp ( - r/\lambda )}}{r},
\end{eqnarray}%
where $Z_i$ is the charge number of the particle of the $i$-th specie, $e=4.8032 \times 10^{-10}{(\rm{ESU})}$ is the charge of an electron and $\lambda$ is the larger of Debye length $\lambda _D$ or the average interionic distance $\lambda _0$:
\begin{eqnarray}
&&\lambda  = \max \{ {\lambda _D},{\lambda _0}\} , \nonumber \\
&&{\lambda _D} = \sqrt {\frac{{kT}}{{4\pi {e^2}\sum\limits_{\textrm{ions+electron}} {{n_i}{Z_i}^2} }}} , \nonumber \\
&&{\lambda _0} = \sqrt[3]{{\frac{3}{{4\pi \sum\limits_{\textrm{ions}} {{n_i}} }}}},
\end{eqnarray}%

By defining a dimensionless distance $R=r/\lambda$, a dimensionless impact parameter $B=b/\lambda$ and a dimensionless couple parameter $\Lambda  = (4kT\lambda )/({Z_i}{Z_j}{e^2})$, the above integrals can be rewritten as dimensionless forms as follows:
\begin{eqnarray}
&&{F^{(ml)}} (\Lambda ) ={[\pi {(\frac{{{Z_i}{Z_j}{e^2}}}{{2kT}})^2}{\sqrt{\frac{{kT (m_i+m_j) }}{{2\pi {m_i m_j}}}}}]^{ - 1}}{\Omega _{ij}}^{(ml)} \nonumber \\
&&= \frac{{{\Lambda ^2}}}{8}\int\limits_0^{ + \infty } {{{\textrm{exp}}(- X)}{X^{l + 1}}{\Phi ^{(m)}}(\Lambda X)dX} , \label{F_0}
\end{eqnarray}%
\begin{eqnarray}
&&{\Phi ^{(m)}}(a) = \frac{{{\phi ^{(m)}}}}{{2\pi {\lambda ^2}}} = \int\limits_0^{ + \infty } {[1 - {{\cos }^m}\chi (a,{R_0})]d{B^2}} , \label{Phi_0}
\end{eqnarray}%
\begin{eqnarray}
&&\chi (a,{R_0}) = \pi  - \int\limits_{R_0}^{ + \infty } { \frac {2BdR} { {R}\sqrt {R^2 - B^2 - 4R \exp ( - R)/a}} }. \label{chi_0}
\end{eqnarray}%
where the relation among $a$, $B$ and $R_0(=r_{0,ij}/\lambda)$ is
\begin{eqnarray}
&&{B^2}={R_0}^2 - \frac{4}{a}{R_0}\exp ( - {R_0}). \label{R_0}
\end{eqnarray}%

\section{Numerical integral} \label{numintegral}

\subsection{transform of integral $\Phi$}

The dimensionless collision cross sections $\Phi$ is determined by Eq.(\ref{Phi_0}). It is better to change the independent variable from $B$ to $R_0$ than directly integration over $B$ in order to avoid solving $R_0$ from Eq.(\ref{R_0}) \citep{ppfm86}. On another hand, for very small value of $\chi$, the truncation error of numerical subtraction $1-cos^m\chi$ may be relatively large. Therefore I integrate $\Phi$ for $m=1$ and $m=2$ by using this form:
\begin{eqnarray}
&&{\Phi ^{(3-m)}}(a) =m \int\limits_{B = 0}^{ + \infty } {{\sin }^2}\frac{{\chi (a,{R_0})}}{m}
\frac{\partial B^2}{\partial R_0}d{R_0},
\end{eqnarray}%
where $\partial B^2 / \partial R_0 = 2{R_0} + 4({R_0} - 1)\exp ( - {R_0})/a$.

\begin{figure}
\includegraphics[scale=0.6]{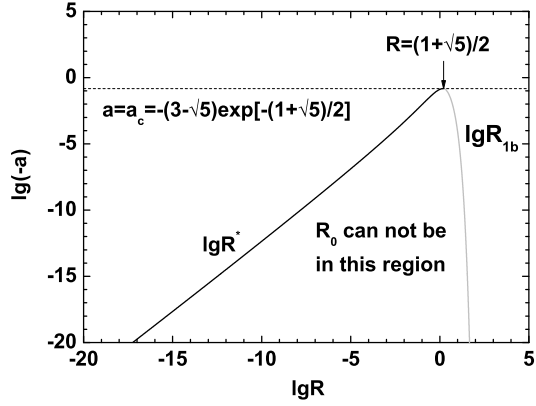}
\caption{ The limits $R^*$ and $R_{1b}$ of integral $\Phi$ as functions of $a$. \label{FigR} }
\end{figure}

In order to obtain the interval of $R_0$ for the integral, it is necessary to investigate the behaviors of the function $B(a,{R_0})$. This has been done and the details are shown in Appendix \ref{appendix1}. It is shown that, for $a_c =- (3 - \sqrt 5 )\exp [ - (1 + \sqrt 5 )/2]  <a<0$ and an arbitrary $B\geq0$, $R_0$ can not locate in $R^*(a) < R_0 < R_{1b}(a)$, where $R_{1b}$ is the larger root of $\partial B^2(a,{R_0}) / \partial R_0 = 0 $ and $R^*$ is determined by $B^2(a,{R^*})=B^2(a,{R_{1b}})$ and $R^*<R_{1b}$. Figure \ref{FigR} shows $R^*$ and $R_{1b}$ as functions of $a$. The existing of the forbidden region of $R_0$ is caused by more than one radicand in Eq.(\ref{R_0}), see also \citet{bf14}.

Accordingly, for repulsive potential ($a>0$):
\begin{eqnarray}
&&{\Phi ^{(3-m)}}(a) = m \int\limits_{R_0\mid_{B=0}}^{ + \infty } {{\sin }^2}\frac{{\chi }}{m} \frac{\partial B^2}{\partial R_0} d{R_0}, \label{Fi_p}
\end{eqnarray}%
for attractive potential ($a<0$) with the parameter $a\leq a_c$:
\begin{eqnarray}
&&{\Phi ^{(3-m)}}(a) = m \int\limits_{0}^{ + \infty } {{\sin }^2}\frac{{\chi }}{m} \frac{\partial B^2}{\partial R_0}d{R_0}, \label{Fi_a1}
\end{eqnarray}%
and for attractive potential with the parameter $a_c<a<0$:
\begin{eqnarray}
&&{\Phi ^{(3-m)}}(a) = m ( \int\limits_0^{{R^*(a)}} + \int\limits_{{R_{1b}(a)}}^{ + \infty }) {{\sin }^2}\frac{{\chi }}{m} \frac{\partial B^2}{\partial R_0} d{R_0}. \label{Fi_a2}
\end{eqnarray}%

\subsection{Numerical calculations of integral $\chi$}

The scattering angle $\chi$ is a basic integral in the calculation of the integral $F$. Therefore it is necessary to calculate $\chi$ in a high accuracy in order to ensure the accuracy of $F$. By using a change of variable:
\begin{eqnarray}
&&z = \sqrt {1 - \frac{{{R_0}}}{R}},
\end{eqnarray}%
the improper integral $\chi$, e.g., Eq.(\ref{chi_0}), can be rewritten as a normal finite integral
\begin{eqnarray}
&&\chi (a,{R_0}) = \pi  - 4\int\limits_0^1 {\frac{1}{{\sqrt {\alpha  + 2 - {z^2}} }}} dz, \label{chi_1}
\end{eqnarray}%
where $\alpha$ is defined by:
\begin{eqnarray}
&&\alpha  = \frac{u}{{{z^2}}}[1 - (1 - {z^2})\exp ( - \frac{{{z^2}}}{{1 - {z^2}}}{R_0})]
\end{eqnarray}%
and
\begin{eqnarray}
&&u = \frac{4}{a}\frac{{{R_0}\exp ( - {R_0})}}{{{B^2}(a,{R_0})}} = \frac{4}{{a{R_0}\exp ({R_0}) - 4}}. \label{Def-u}
\end{eqnarray}%

Directly calculation the integral by using the form Eq.(\ref{chi_1}) could lead to a problem. For large parameter $(a,R_0)$, $\alpha$ is small. As $\alpha$ becomes very smalle, $\chi$ should be close to zero. The numerical subtraction in Eq.(\ref{chi_1}) can result in significant relative error of $\chi$. In order to avoid that problem, I rewritten Eq.(\ref{chi_1}) as follow:
\begin{eqnarray}
&&\chi = 4\int\limits_0^1 {\frac{1 }{{\sqrt {\alpha  + 2 - {z^2}} \sqrt {2 - {z^2}} }}} \times \nonumber \\
&&{\frac{\alpha }{{\sqrt {\alpha  + 2 - {z^2}}  + \sqrt {2 - {z^2}}  }}} dz.
\end{eqnarray}%

It is necessary to investigate the properties of the integrand for improving the accuracy of numerical integral. If the value of integrand is very high in a small interval $(z_1,z_2)$, it should be ensured in the numerical calculation of the integral that the resolution in the interval is enough. Therefore the parameter ranges of $(a,R_0)$ around the singularities of the integrand should be seriously concerned.

For repulsive potential, $a>0$, thus $\alpha>0$. There is no singularity of integrand. The integrand is a slowly varying function.

For attractive potential, $a<0$, if $z=0$ is a singularity, $\alpha(z=0)$ should be equal to $-2$. As $z \rightarrow 0$, the behavior of $\alpha$ is
\begin{eqnarray}
&&\alpha  = u\frac{{1 - (1 - {z^2})  \exp(-\frac{z^2}{1 - {z^2}}R_0)  }}{{{z^2}}} \rightarrow \nonumber \\
&&- u{\{ \frac{\partial }{{\partial x}}[(1 - x) \exp(-\frac{{x}}{1 - x}R_0) ]\} _{x = z^2 = 0}} \nonumber \\
&&= u(1 + {R_0}).
\end{eqnarray}%
According to Eq.(\ref{Def-u}) and the above equation, $\alpha(z=0)=-2$ is equivalent to
\begin{eqnarray}
&&2{R_0} + \frac{4}{a}({R_0} - 1)\exp ( - {R_0}) = 0.
\end{eqnarray}%
According to Appendix \ref{appendix1}, this occurs only if $a>a_c$ and $R_0=R_{1b}$.

For attractive potential ($a<0$), if $0<z_0<1$ is a singularity, $\alpha(z=z_0)$ should be equal to $z_0^2-2$, which is equivalent to
\begin{eqnarray}
&&{R^2} - \frac{4}{a}R\exp ( - R) - {B^2} = 0,
\end{eqnarray}%
where $R=R_0/(1-z_0^2)$. According to Appendix \ref{appendix1}, this occurs only if $a_c<a<0$, $R_0=R^*$ and $R=R_{1b}$. The singularity is $z=\sqrt{1-R_0/R_{1b}}$.

For attractive potential ($a<0$), if $z=1$ is a singularity, $\alpha(z=1)$ should be equal to $-1$. As $z \rightarrow 1$, the behavior of $\alpha$ is
\begin{eqnarray}
&&\alpha  = \frac{u}{{{z^2}}}[1 - (1 - {z^2})\exp ( - \frac{{{z^2}}}{{1 - {z^2}}}{R_0})] \rightarrow u.
\end{eqnarray}%
Therefore $z=1$ is a singularity only if $a \rightarrow 0^-$ or $R_0=0$.

In all singularity cases, the behavior of integrand near a singularity $z_0$ is $\sim1/(z-z_0)$, thus $\chi$ is divergent. However, for the integral $\Phi$, that is not a problem since the integrand of $\Phi$ is $1-cos^m \chi $ which is always limited whatever $\chi$ is infinite or not. The contribution of the singularities of $\chi$ to $\Phi$ is zero.

\begin{figure}
\centering
\includegraphics[scale=0.6]{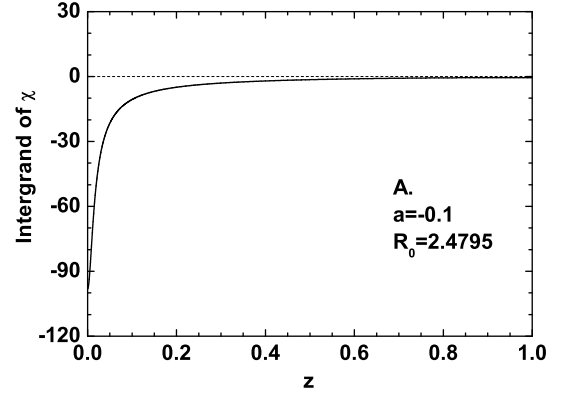}
\includegraphics[scale=0.6]{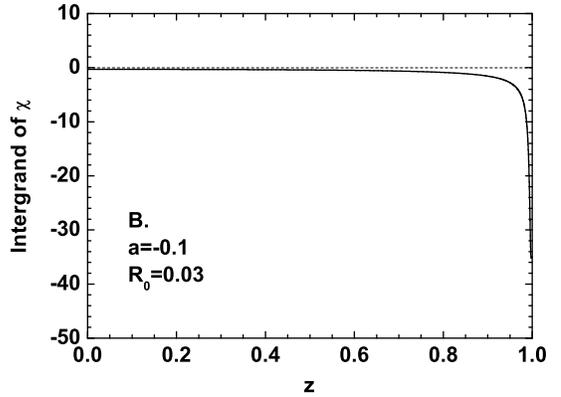}
\includegraphics[scale=0.6]{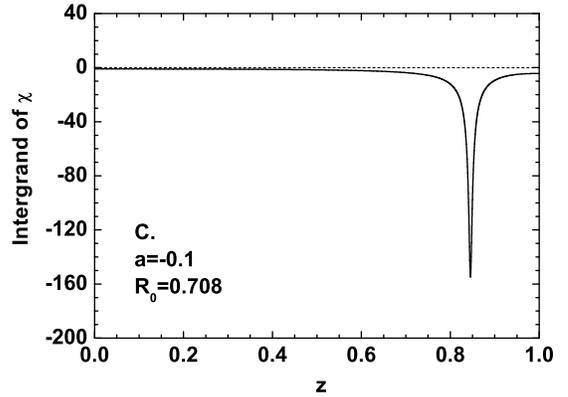}
\caption{ The integrand of $\chi$ as a function of $z$ for different $a$ and $R_0$. When $a=-0.1$, $R_*\approx 0.70873$ and $R_{1b} \approx 2.47933$. According to the analysis (see text), $z=0$ is a singularity if $R_0=R_{1b}$, $z=1$ is a singularity if $R_0=0$ and $z=\sqrt{1-R^*/R_{1b}} \approx 0.8451$ is a singularity if $R_0=R_*$. \label{Xcore} }
\end{figure}

When the values of parameters $(a, R_0)$ are close to the conditions of singularities, near the corresponding singularity, the value of the integrand of $\chi$ may be very high and a small subinterval may be the main contributor of $\chi$. This must be concerned in the numerical calculation of $\chi$. Fig.\ref{Xcore} shows the value of the integrand of $\chi$ in three cases. It is shown that, as the parameter $R_0$ close to the singularity conditions, the value of the integrand of $\chi$ may quickly change near the corresponding singularity. It is obvious that, in those cases, directly using a low-order Gauss-Legendre formula to calculate $\chi$ from $z=0$ to $z=1$ may obtain an incorrect result.

In the numerical calculation of $\chi$, since there are three possible singularities in the integrand of $\chi$, I split the integral into subintervals with the boundaries being the possible singularities as follows.
For $a>0$ or $a<a_c$, $\chi$ is the sum of two integrals (from $z_a=0$ to $z_b=0.9$ and $z_a=1$ to $z_b=0.9$). For $a_c<a<0$, $\chi$ is the sum of four integrals (from $z_a=0$ to $z_b=0.5\sqrt{1-R_0/R_{1b}}$, $z_a=\sqrt{1-R_0/R_{1b}}$ to $z_b=0.5\sqrt{1-R_0/R_{1b}}$, $z_a=\sqrt{1-R_0/R_{1b}}$ to $z_b=(1+\sqrt{1-R_0/R_{1b}})/2$ and $z_a=1$ to $z_b=(1+\sqrt{1-R_0/R_{1b}})/2$). In order to improve the numerical accuracy of $1-z^2$, I trace $z'=1-z$ in the calculations. When $R_0 / R_{1b}<10^{-4}$, $1-\sqrt{1-R_0/R_{1b}}$ is calculated by using its power series from $(R_0/R_{1b})^1$ to $(R_0/R_{1b})^6$ terms.
The details of the calculation of the integral from $z=z_a$ to $z=z_b$ is as follows.

1: set the initial $z$ as $z_1=z_a$, initial $z'$ as $z'_1=1-z_a$ and an initial space step $d$:
for $z_a=0$,
\begin{eqnarray}
&&d = \min \{ d{z_{\max }},\sqrt {2\varepsilon\frac{{(1 + {R_0})u + 2}}{{u{R_0}^2 + 2}}} ,\frac{\varepsilon}{{\sqrt {{R_0} + 1} }}\},
\end{eqnarray}%
for $z_a=1$,
\begin{eqnarray}
&&d = \min \{ d{z_{\max }},2\varepsilon,\frac{{0.5{R_0}}}{{\max (100{R_0},100)}} \},
\end{eqnarray}%
and for $z_a=\sqrt{1-R_0/R_{1b}}$,
\begin{eqnarray}
&&d = \min \{ d{z_{\max }},\varepsilon {\left. {\left| {\frac{{\partial z}}{{\partial \ln (\alpha  + 2 - {z^2})}}} \right|} \right|_{{z_a}}}\} ,
\end{eqnarray}%
 where $\varepsilon=0.001$, $ d_{\rm{max}} = 0.01\lambda_{\chi}$ and $\lambda_{\chi}$ is a parameter which controls the accuracy of the numerical calculation of $\chi$. The above formulas of initial space step $d$ are based on analysis of the behaviors of $\alpha  + 2 - {z^2}$ and its derivative respect to $z$. The aim is to ensure that, in the initial space step $d$, $\alpha  + 2 - {z^2}$ does not change too much.

2: set $z_2=z_1 \pm d$, $z'_2=z'_1 \pm d$ ($+$ for $z_b>z_a$ and $-$ for $z_b<z_a$) and calculate the values of the integrand on $z_1$ and $z_2$.

3: check the relative variation of integrand. If the relative difference between the values of the integrand on $z_1$ and on $z_2$ is larger than $ 0.1\lambda_{\chi}$, the space step $ d $ is shorten as $ 0.618d \Rightarrow d  $ and go back to step 2, until the relative variation is less than $ 0.1\lambda_{\chi}$.

4: in subinterval $[{\rm{min}}(z_1,z_2),{\rm{max}}(z_1,z_2)]$, use 64-point Gauss-Legendre formula to evaluate integral.

5: update $z_1$ as $z_2 \Rightarrow z_1$, $z'_1$ as $z'_2 \Rightarrow z'_1$ and adjust the space step $d$ as $ {\rm{min}} \{ (1+0.2\lambda_{\chi})d, d_{\rm{max}}, \left| z_2-z_a \right| \} \Rightarrow d $.

6: stop the loop if $z_2=z_b$ and the sum of integrals in all subintervals is the result of $\chi(a,R_0)$, otherwise go back to step 2.

In the numerical calculations of $\chi(a,R_0)$, $z_a$, $z_b$, $z$, $z_1$, $z_2$, $z'$, $z'_1$, $z'_2$ are in 32-digital, other variables are in 16-digital.
The value of the term $\alpha  + 2 - {z^2}$ in the integrand is calculated by using the following form in order to improve the numerical accuracy:
\begin{eqnarray}
&& \alpha  + 2 - {z^2} =  {R_0} \{\frac{{\exp ( - x)}}{{{R_0} + x}} \nonumber \\
&& + [\frac{{1 - \exp ( - x)}}{x}  \frac{{x + 4\exp ( - {R_0}) / a}}{{x + {R_0}}}{R_0} + 1]\frac{{{R_0}}}{{{B^2}}}\} ,
\end{eqnarray}%
where $x=z^2 R_0/[(1+z)z']$ and the value of $[1-{\rm{exp}}(-x)]/x$ is calculated by using power series (from $x^1$ to $x^{20}$ terms) if $x<0.01$.

\subsection{Numerical calculations of integral $\Phi$}

The formula of integral $\Phi(a)$ is Eq.(\ref{Fi_p}), Eq.(\ref{Fi_a1}) or Eq.(\ref{Fi_a2}), determined by the value of $a$. In order to ensure the accuracy of integral $\Phi(a)$, I split the integral into many subintervals as similar as in the calculation of $\chi(a,R_0)$. It is necessary to analyze the behavior of the integrand of $\Phi(a)$, which may help to set proper widths of subintervals. Because $\alpha \propto u$, it is found that $\chi \propto u$. According to the formula of integral $\Phi(a)$, if $\left| u \right| \ll 1$, the integrand is small thus the contribution to $\Phi(a)$ is small. Because
\begin{eqnarray}
&&\left| u \right| = \frac{4}{{\left| {a{R_0}\exp ({R_0}) - 4} \right|}} \ll 1 \nonumber \\
&&\Leftarrow \left| a \right|{R_0}\exp ({R_0}) \gg 1 \Leftarrow {R_0} \gg \frac{1}{{\left| a \right|}},
\end{eqnarray}%
the region $(R_0<1/\left| a \right|)$ may mainly contribute to $\Phi(a)$. When $\left| a \right| \gg 1$, the width of this region is very small. It should be ensure that the resolution is enough in that region.

The details of the numerical calculation of $\Phi(a)$ for Eq.(\ref{Fi_p}), Eq.(\ref{Fi_a1}) and the integral in $[R_{1b},+\infty]$ in  Eq.(\ref{Fi_a2}) are as follows:

1: set $R_1$ to be the low limit of the integral and an initial step $\triangle R_0 =  10^{-10}\lambda_{\Phi}/{\rm{max}}(1,\left| a \right|)$, where $\lambda_{\Phi}$ is a parameter to control the accuracy of integral $\Phi$.

2: let $R_2=R_1+ \triangle R_0$. If $B(a,R_2)<100$, use 64-point Gauss-Legendre formula to evaluate integral in the subinterval $[R_1,R_2]$. If $B(a,R_2)>100$, use 64-point Gauss-Laguerre formula to evaluate integral in the subinterval $[R_1,+\infty)$ and the calculation is finished.

3: if the integral in the subinterval is larger than $0.01\lambda_{\Phi}\Phi_{0}(a)$ where $\Phi_{0}(a)$ is an estimation of $\Phi(a)$, reduce the step as $0.618 \triangle R \Rightarrow  \triangle R$.

4: update $R_1$ as $R_2 \Rightarrow R_1$ and adjust the space step $\triangle R$ as $ {\rm{min}} \{ (1+0.2\lambda_{\Phi})\triangle R, \triangle R_{\rm{max}} \} \Rightarrow \triangle R$ where $\triangle R_{\rm{max}} = 10 \lambda_{\Phi}$, go to step 2.

The effect of step 3 is to ensure that the space step $\triangle R$ is small enough when the integrand is larger, which ensure the accuracy of the numerical result of $\Phi(a)$.
The initial estimation of $\Phi(a)$ is obtained by using the above loops except step 3. The adopted estimation of $\Phi(a)$, e.g., $\Phi_{0}(a)$, is obtained by using the above loops and the initial estimation of $\Phi(a)$. The difference between $\Phi_{0}(a)$ and the final result of $\Phi(a)$ is very small.

The details of the numerical calculation of $\Phi(a)$ for the integral in $[0,R^*]$ in Eq.(\ref{Fi_a2}) are similar to the above loops. But the integral is firstly split to two part, e.g., in $[0,0.5R^*]$ and in $[0.5R^*,R^*]$. For the $[0.5R^*,R^*]$ part, it is started from $R_0=R^*$ and using negative steps $\triangle R$.

I have calculated 8193 data points of $\Phi^{(m)}(a)$ for $m=1$ and $m=2$ in the range of $-20 \leq lg \left| a \right| \leq 40$ with the step of $lg \left| a \right|$ as $60/8192$.
There are two parameters controlling the accuracy of numerical integral $\Phi(a)$: $\lambda_{\chi}$ and $\lambda_{\Phi}$. I have tested three cases, e.g., $(\lambda_{\chi}=5,\lambda_{\Phi}=1)$, $(\lambda_{\chi}=5,\lambda_{\Phi}=2)$ and $(\lambda_{\chi}=10,\lambda_{\Phi}=1)$, to investigate the variations of resulting $\Phi(a)$. It is found that the maximum of relative differences is less than $10^{-14}$. Therefore the relative errors of $\Phi(a)$ calculated by using $(\lambda_{\chi}=5,\lambda_{\Phi}=1)$ should be in that level.

\subsection{Numerical calculations of integral $F$}

The formula of integral $F$ is Eq.(\ref{F_0}). It can be rewritten as
\begin{eqnarray}
&&{F^{(ml)}}(\Lambda ) = \frac{{{{\left| \Lambda  \right|}^{ - l}}}}{8}\int\limits_0^{ + \infty } {{{\textrm{exp}}{( - \frac{a}{{\left| \Lambda  \right|}})}}{a^{l + 1}}{\phi ^{(m)}}(\frac{\Lambda }{{\left| \Lambda  \right|}}a)da}.
\end{eqnarray}%

The values of $\Phi^{(m)}(x)$ for $-20 \leq lg \left| x \right| \leq 40$ with the step of $lg \left| x \right|$ as $60/8192$ have been calculated. I use the numerical results of integral $F$ in $-10^{-20} \leq a \leq 10^{40}$ to be the final result of $F$, namely, the integral $F$ in intervals $a < 10^{-20}$ and $a>10^{40}$ are ignored. Let me estimate the integral of $F$ in intervals $a < 10^{-20}$ and $a>10^{40}$ to investigate whether they are ignorable.

By using the numerical results of $\Phi^{(m)}(x)$ in $-20 \le \lg \left| x \right| \le -15$, it is found that ${\phi ^{(m)}}(x) \approx {{{{10}^{2.3}}}}/({{{\sqrt{m} \left| x \right|  ^{0.047}}}})$ is an approximation in $\sim15\%$. I assume that the approximation holds for $\lg \left| x \right| \leq -15$. In this case, if $\left| \Lambda  \right| \gg {10^{-20}}$, integral of $F$ in interval $a<10^{-20}$ can be estimated as
\begin{eqnarray}
&&\frac{{{{\left| \Lambda  \right|}^{ - l}}}}{8}\int\limits_0^{{{10}^{ - 20}}} {{{\textrm{exp}}{( - \frac{a}{{\left| \Lambda  \right|}})}}{a^{l + 1}}{\phi ^{(m)}}(\frac{\Lambda }{{\left| \Lambda  \right|}}a)da} \nonumber \\
&&\approx \frac{{{{\left| \Lambda  \right|}^{ - l}}}}{8}\frac{{{{10}^{2.3}}}}{{\sqrt m }}\int\limits_0^{{{10}^{ - 20}}} {{{\textrm{exp}}{( - \frac{a}{{\left| \Lambda  \right|}})}}{a^{l + 0.953}}da} \nonumber \\
&&< \frac{{{{\left| \Lambda  \right|}^{ - l}}}}{8}\frac{{{{10}^{2.3}}}}{{\sqrt m }}{10^{ - 20}}^{(l + 1.953)} < {10^{ - 37}}.
\end{eqnarray}%

By using the numerical results of $\Phi^{(m)}(x)$ in $30 \le \lg \left| x \right| \le 40$, it is found that ${\phi ^{(m)}}(x) \approx {{{{10}^3}m}}/{{{x^2}}}$ is an approximation in $\sim 50\%$. I assume that the approximation holds for $\lg \left| x \right| \geq 30$. In this case, if $\left| \Lambda  \right| \ll {10^{40}}$, integral of $F$ in interval $a>10^{40}$ can be estimated as
\begin{eqnarray}
&&\frac{{{{\left| \Lambda  \right|}^{ - l}}}}{8}\int\limits_{{{10}^{40}}}^{ + \infty } {{{\textrm{exp}}{( - \frac{a}{{\left| \Lambda  \right|}})}}{a^{l + 1}}{\phi ^{(m)}}(\frac{\Lambda }{{\left| \Lambda  \right|}}a)da} \nonumber \\
&&\approx \frac{{{{\left| \Lambda  \right|}^{ - l}}}}{8}\int\limits_{{{10}^{40}}}^{ + \infty } {{{\textrm{exp}}{( - \frac{a}{{\left| \Lambda  \right|}})}}{a^{l + 1}}\frac{{{{10}^3}m}}{{{a^2}}}da} \nonumber \\
&&= \frac{{{{10}^3}m}}{8}\int\limits_{\frac{{{{10}^{40}}}}{{\left| \Lambda  \right|}}}^{ + \infty } {{{\textrm{exp}}{( - x)}}{x^{l - 1}}dx} \nonumber \\
&&= \frac{{{{10}^3}m}}{8}(l - 1)!{{\textrm{exp}}{( - \frac{{{{10}^{40}}}}{{\left| \Lambda  \right|}})}}\sum\limits_{k = 0}^{l - 1} {\frac{1}{{k!}}} {(\frac{{{{10}^{40}}}}{{\left| \Lambda  \right|}})^k} \nonumber \\
&&\approx \frac{{{{10}^3}m}}{8}{{\textrm{exp}}{( - \frac{{{{10}^{40}}}}{{\left| \Lambda  \right|}})}}{(\frac{{{{10}^{40}}}}{{\left| \Lambda  \right|}})^{l - 1}} \nonumber \\
&& \le 250{{\textrm{exp}}{( - \frac{{{{10}^{40}}}}{{\left| \Lambda \right|}})}}{(\frac{{{{10}^{40}}}}{{\left| \Lambda \right|}})^2}.
\end{eqnarray}%
The final term increases quickly as $\psi$ ($={\rm{ln}}[{\rm{ln}}(\Lambda^2+1)]$) increasing. For $\psi$ being less than $5.16$, $5.17$, $5.18$, $5.19$ and $5.20$, the value is less than $10^{-59}$, $10^{-21}$, $10^{-6}$, $10^0$ and $10^2$.

Comparing with the numerical results of integral $F$ in $-10^{-20} \leq a \leq 10^{40}$ for $-18 \leq \psi \leq 5.17$, the integral of $F$ in intervals $a < 10^{-20}$ and $a>10^{40}$ are actually ignorable because the ratios are less than $10^{-20}$.

In the range $-18 \leq \psi \leq 5.17$, the numerical integrals of $F^{(ml)}$ in $10^{-20} \leq a \leq 10^{40}$ are calculated by using Romberg's method (repeatedly applying Richardson extrapolation on the trapezium rule) in order to obtain results with high accuracy. The estimate relative errors of $F^{(ml)}$, which can be obtained in the last two iterations in Romberg's method, are $<10^{-12}$ for attractive potential ($\Lambda<0$) and $<10^{-14}$ for repulsive potential ($\Lambda>0$).

\section{Results} \label{results}

By using the numerical calculation scheme introduced in the above section, the integral $F^{(11)}$, $F^{(12)}$, $F^{(13)}$ and $F^{(22)}$ can be calculated in the range $-18 \leq \psi={\rm{ln}}[{\rm{ln}}(\Lambda^2+1)] \leq 5.17$ for both repulsive potential ($\Lambda>0$) and attractive potential ($\Lambda<0$) with the accuracies $<10^{-12}$ for attractive potential ($\Lambda<0$) and $<10^{-14}$ for repulsive potential ($\Lambda>0$). The screened Coulomb potential is invalid for strong coupled plasmas. The errors of collision integrals become significant for $\Gamma > 10$ \citep[see e.g.,][Fig.7]{bd14}, which corresponding to about $ \psi < -5 $. In this paper, I show the 121 data of $F$ for $-7 \leq \psi \leq 5$, which covers the range studied by \citet{ppfm86}, with the step of $\psi$ as $0.1$ in the tables in Appendix \ref{appendix2}. This should be enough for normal applications in stellar evolution. In the tables, the first and second order derivatives $d {\rm{ln}} F^{(ml)} / d \psi$ and $d^2 {\rm{ln}} F^{(ml)} / d \psi^2$ are also shown in the tables. The estimate errors of the derivatives are less than $ \sim 10^{-11}$. By using the data in the tables, one can use Hermite fifth order polynomial interpolation to calculate $F^{(ml)}$ in each interval of $\psi$:
\begin{eqnarray}
&&Y = (1 + 3X + 6{X^2}){(1 - X)^3}{Y_k} \nonumber \\
&&+ X(1 + 3X){(1 - X)^3}h{Y_k}' \nonumber \\
&&+ \frac{1}{2}{X^2}{(1 - X)^3}{h^2}{Y_k}'' \nonumber \\
&&+[1 + 3(1 - X) + 6{(1 - X)^2}]{X^3}{Y_{k + 1}} \nonumber \\
&&-[(1 - X) + 3{(1 - X)^2}]{X^3}h{Y_{k + 1}}' \nonumber \\
&&+ \frac{1}{2}{(1 - X)^2}{X^3}{h^2}{Y_{k + 1}}'',
\end{eqnarray}%
where $0 \le X = (\psi  - {\psi _k})/h \le 1$, $h={\psi _{k + 1}} - {\psi _k}=0.1$
and $Y$ can be each ${\rm{ln}} F^{(ml)}$.
The differences of ${\rm{ln}}F$ between using the Hermite fifth order polynomial interpolation and the numerical integral results of ${\rm{ln}}F$ are less than $ \sim 10^{-10}$. Even using the Hermite cubic polynomial interpolation, the differences are less than $ \sim 10^{-6}$.

In the case of very weak coupled plasma (i.e., $\psi>4.5$), all $F^{(ml)}$ are essentially proportional to ${\rm{exp}}(\psi)$ and the differences of $F^{(ml)}$ between the repulsive potential case and the attractive potential case are very small \citep{ppfm86}.
Therefore the following simple formulas, which are the best fittings of the numerical results of $F^{(ml)}$ in $4.5<\psi<5.17$ with relative errors $ < 10^{-11}$, should hold for $\psi>4.5$:
\begin{eqnarray}
&&F^{(11)} = {\rm{exp}}(\psi)-3.3088626596, \label{F1_fit} \\
&&F^{(12)} = F^{(11)}+2, \label{F2_fit} \\
&&F^{(13)} = 2{\rm{exp}}(\psi)-0.6177253192, \label{F3_fit} \\
&&F^{(22)} = F^{(13)}-4. \label{F4_fit}
\end{eqnarray}%

\section{Comparing with others results} \label{linktoothers}

In Burgers equation, there are resistance coefficients $K_{ij}$, $z_{ij}$, $z_{ij}'$ and $z_{ij}''$. They are determined by the integral $F$ as follows \citep{m84,ppfm86}:
\begin{eqnarray}
&&K_{ij}=\frac{2}{3}{n_i}{n_j}{(\frac{{{Z_i}{Z_j}{e^2}}}{{kT}})^2}\sqrt {2\pi kT \frac{{{m_i}{m_j}}}{{{m_i} + {m_j}}}} F_{ij}^{(11)}, \\
&&z_{ij}=1-\frac{2F_{ij}^{(12)}}{5F_{ij}^{(11)}}, \\
&&z_{ij}'=\frac{5}{2}-\frac{2}{5}\frac{5F_{ij}^{(12)}-F^{(13)}}{F_{ij}^{(11)}}, \\
&&z_{ij}''=\frac{F_{ij}^{(22)}}{F_{ij}^{(11)}},
\end{eqnarray}%
where
\begin{eqnarray}
&&F_{ij}^{(ml)}=F^{(ml)}(\Lambda_{ij}), \\
&&\Lambda_{ij}=\frac{4kT\lambda}{{Z_i}{Z_j}{e^2}},
\end{eqnarray}%
and $n_i$ is the number density of the $i$-th species.

In the truncated pure Coulomb potential case, the resistance coefficients are \citep{m84,ppfm86}:
\begin{eqnarray}
&&K_{ij}^{(0)}=\frac{2}{3}{n_i}{n_j}{(\frac{{{Z_i}{Z_j}{e^2}}}{{kT}})^2}\sqrt {2\pi kT \frac{{{m_i}{m_j}}}{{{m_i} + {m_j}}}} \times \nonumber \\
&&{\rm{ln}}[1+(\frac{4kT\lambda}{{Z_i}{Z_j}{e^2}})^2], \\
&&z_{ij}^{(0)}=0.6, \\
&&z_{ij}'^{(0)}=1.3, \\
&&z_{ij}''^{(0)}=2.
\end{eqnarray}%

\begin{figure*}
\centering
\includegraphics[scale=0.6]{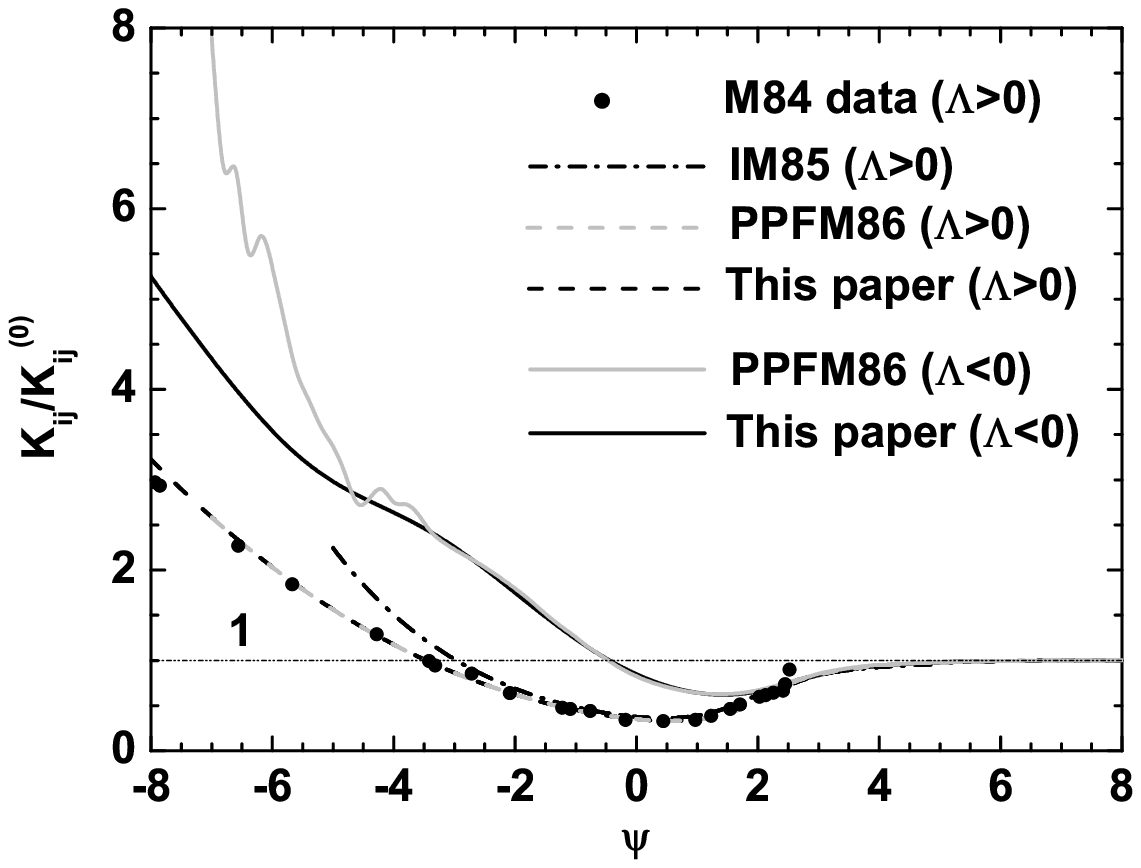}
\includegraphics[scale=0.6]{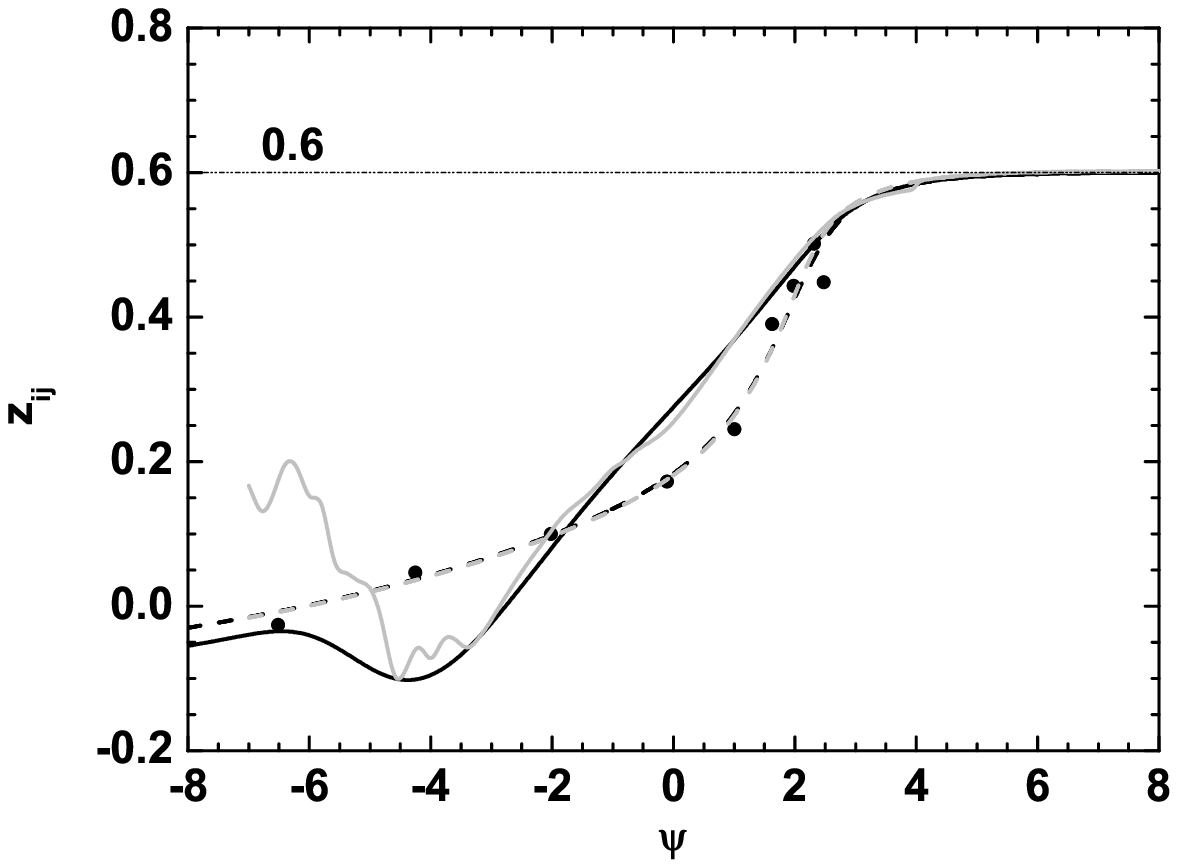}
\includegraphics[scale=0.6]{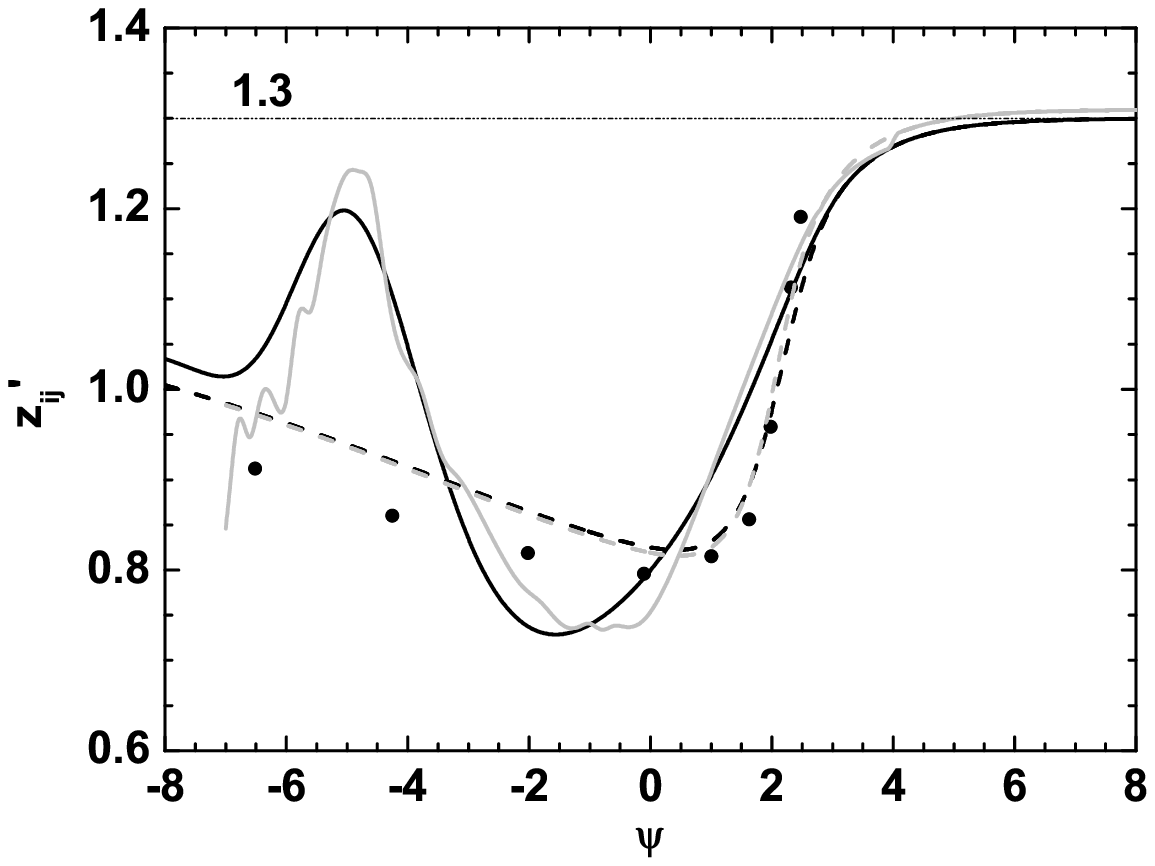}
\includegraphics[scale=0.6]{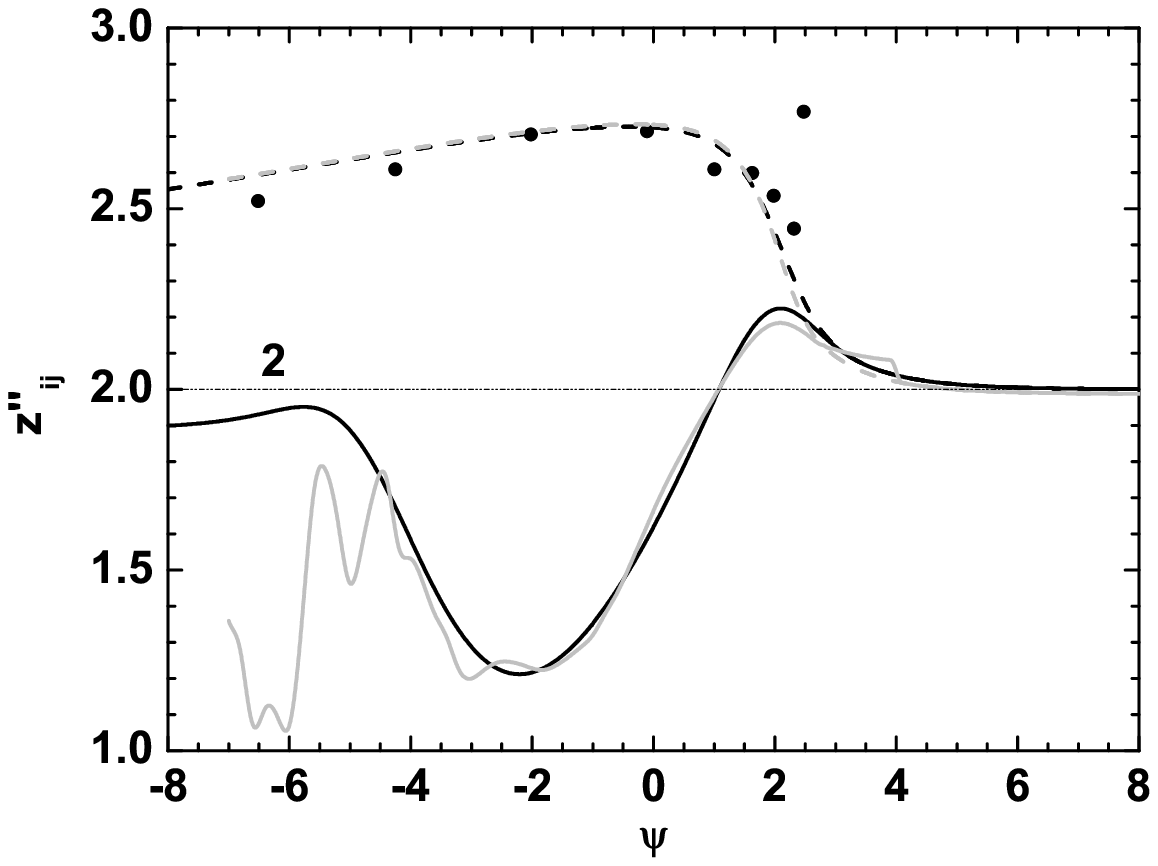}
\caption{ The resistance coefficients computed by \citet[][M84]{m84}, \citet[][IM85]{im85} fitting  on \citeauthor{fm79}'s (\citeyear{fm79}) results, \citet[][PPFM86]{ppfm86} and this paper. \label{KZvs}}
\end{figure*}

The comparison of resistance coefficients among different works, e.g., \citet[][M84]{m84}, \citet[][IM85]{im85}, \citet[][PPFM86]{ppfm86} and this paper are shown in Fig.\ref{KZvs}. M84 and IM85 have not computed the attractive potential case. It is shown that, for the repulsive potential, PPFM86's results are very close to this paper, M84's results show divergence for larger $\psi$ and IM85's fitting is a good approximation in $\psi>-2$, i.e., the plasma is not strongly coupled. For the attractive potential, the differences between PPFM86's results and this paper are significant in intermediate and strong coupled case (i.e., $\psi<0$) in which PPFM86's results show oscillatory behaviors. In the weak coupled limit (large $\psi$), PPFM86's results do not strictly reproduce the truncated pure Coulomb potential results. According to their analytical fitting,  i.e., Eq.(73a-73d) in their paper, in the weak screening limit case, $K_{ij}/K_{ij}^{(0)}=1.0014$, $z_{ij}=0.6023$, $z'_{ij}=1.3098$ and $z''_{ij}=1.9874$. In the weak coupled limit case, the fitting in this paper (Eqs.\ref{F1_fit}-\ref{F4_fit}) show exactly the same results to the truncated pure Coulomb potential case.

\section{Applications} \label{app}

\subsection{the diffusion velocity: the solution of Burgers equations}

The diffusion coefficients of transport of stellar plasma can be evaluated by using Chapman-Enskog theory \citep{cc70} or Burgers equations \citep{bur69} and those two theories are ultimately equivalent to each other in the limit of collision-dominated plasma \citep{bur69}. In this paper, the latter is adopted to calculate the diffusion coefficients. Burgers equations and the equlibrium conditions of current neutrality and local mass conservation are as follows:
\begin{eqnarray}
&&\sum\limits_{j = 1}^I {{Z_j}{n_j}{w_j}}  = 0, \\
&&\sum\limits_{j = 1}^I {{A_j}{n_j}{w_j}}  = 0, \\
&&\frac{{d{P_i}}}{{dr}} + {n_i}({A_i}{m_u}g - {Z_i}e \textsc{E} ) \nonumber \\
&&= \sum\limits_{j = 1}^I {{K_{ij}}[({w_j} - {w_i}) + {z_{ij}}({x_{ij}}{r_i} - {y_{ij}}{r_j})]}, \\
&&\frac{5}{2}{n_i}k\frac{{dT}}{{dr}} =  - \frac{5}{2}\sum\limits_{j = 1}^I {{K_{ij}}{z_{ij}}{x_{ij}}({w_j} - {w_i})}  + \nonumber \\
&&\sum\limits_{j = 1}^I {{K_{ij}}{y_{ij}}{x_{ij}}(3 + z{'_{ij}} - 0.8z{_{ij}}''){r_j}} \nonumber \\
&&- \sum\limits_{j = 1}^I {{K_{ij}}(3{y_{ij}}^2 + {x_{ij}}^2z{'_{ij}} + 0.8{y_{ij}}{x_{ij}}z{_{ij}}''){r_i}},
\end{eqnarray}%
where $I$ is the number of concerned species, $w_i$, $r_i$, $P_i$ are the diffusion velocity, residual heat flow and pressure of the $i$-th species (for $i<I$, corresponding to ions, and for $i=I$, corresponding to electrons), $A_i=m_i/m_u$ is the mass number of the $i$-th species, $m_i$ is the mass of a single particle of the $i$-th species, $m_u$ is the atomic mass unit, $\textsc{E}$ is the electric field, $g$ is the gravitational acceleration, $x_{ij}=m_j/(m_i+m_j)$ and $y_{ij}=m_i/(m_i+m_j)$.

By defining an unknown vector $s$ as
\begin{eqnarray}
&&s = \left[ {\begin{array}{*{20}{c}}
   w  \\
   r  \\
   {{m_u}g}  \\
   {e \textsc{E} }  \\
\end{array}} \right],
\end{eqnarray}%
two vectors $C$ and $D$ and two matrix $A$ and $B$, the equations can be rewritten as:
\begin{eqnarray}
&&C^T s  = 0, \label{Burgers_charge} \\
&&D^T s  = 0, \label{Burgers_mass} \\
&&\frac{{d[P]}}{{dr}}  = As, \label{Burgers_P} \\
&&k\frac{{dT}}{{dr}} [n] = \frac{2}{5}Bs, \label{Burgers_T}
\end{eqnarray}%
where $[P] = {\left[ {\begin{array}{*{20}{c}}
   {{P_1}} & {{P_2}} & {...} & {{P_{I-1}}} & {{P_{I}}}  \\
\end{array}} \right]^T}$ is the pressure vector and $[n] = {\left[ {\begin{array}{*{20}{c}}
   {{n_1}} & {{n_2}} & {...} & {{n_{I-1}}} & {{n_{I}}}  \\
\end{array}} \right]^T}$ is the number density vector.
The elements of vectors $C$ and $D$ and matrices $A$ and $B$ are as follows:
\begin{eqnarray}
{\textrm{for }}&& 1 \le j \le I: \nonumber \\
&&{C_j} = {Z_j}{n_j}, {D_j} = {A_j}{n_j}, \nonumber \\
{\textrm{for }}&& j > I: \nonumber \\
&&{C_j} = 0, {D_j} = 0, \nonumber \\
{\textrm{for }}&& 1 \le i \le I,1 \le j \le 2I + 2: \nonumber \\
{\textrm{for }}&& i \ne j,1 \le j \le I: \nonumber \\
&&{A_{ij}} = {K_{ij}}, \nonumber \\
&&{A_{i,j + I}} =  - {K_{ij}}{z_{ij}}{y_{ij}}, \nonumber \\
&&{B_{ij}} =  - \frac{5}{2}{K_{ij}}{z_{ij}}{x_{ij}}, \nonumber \\
&&{B_{i,j + I}} = {K_{ij}}{y_{ij}}{x_{ij}}(3 + z{'_{ij}} - 0.8z{''_{ij}}), \nonumber \\
{\textrm{for }}&& 1 \le i \le I: \nonumber \\
&&{A_{ii}} = {K_{ii}} - \sum\limits_{l = 1}^I {{K_{il}}}, \nonumber \\
&&{A_{i,i + I}} =  - {K_{ii}}{z_{ii}}{y_{ii}} + \sum\limits_{l = 1}^I {{K_{il}}{z_{il}}{x_{il}}}, \nonumber \\
&&{B_{ii}} =  - \frac{5}{2}{K_{ii}}{z_{ii}}{x_{ii}} + \frac{5}{2}\sum\limits_{l = 1}^I {{K_{il}}{z_{il}}{x_{il}}}, \nonumber \\
&&{B_{i,i + I}} = {K_{ii}}{y_{ii}}{x_{ii}}(3 + z{'_{ii}} - 0.8z{_{ii}}'')  \nonumber \\
&&- \sum\limits_{l = 1}^I {{K_{il}}(3{y_{il}}^2 + {x_{il}}^2z{'_{il}} + 0.8{y_{il}}{x_{il}}z{_{il}}'')}, \nonumber \\
{\textrm{for }}&& j = 2I + 1: \nonumber \\
&&{A_{ij}} =  - {n_i}{A_i}, \nonumber \\
{\textrm{for }}&& j = 2I + 2: \nonumber \\
&&{A_{ij}} = {n_i}{Z_i},
\end{eqnarray}%

In Burgers equations, the equation of state is assumed as ideal gas, thus:
\begin{eqnarray}
&&\frac{d[P]}{dr} = \frac{d}{dr}(kT[n]) = kT\frac{d[n]}{dr} + k\frac{dT}{dr}[n],
\end{eqnarray}%
and Eqs.(\ref{Burgers_P} \& \ref{Burgers_T}) yield
\begin{eqnarray}
&&kT \frac{d[n]}{dr} = (A-\frac{2}{5}B)s. \label{Burgers_n}
\end{eqnarray}%

Since $dT/dr$ and all $dn_i/dr$ are independent and the equations are linear, the unknown vector $s$ should be a linear combination of $dT/dr$ and $dn_i/dr$. Assume the factors in the linear combination as a vector $\alpha$ and a matrix $\beta$, there is
\begin{eqnarray}
&&s = kT(n \frac{{d\ln T}}{{dr}} \alpha + \beta \frac{d[n]}{dr} ), \label{Dif_S}
\end{eqnarray}%
where $n=\sum\limits_{i = 1}^I {n_i}$ is the total number density of all species.
Take the above equation into Burgers equations, i.e., Eqs.(\ref{Burgers_charge},\ref{Burgers_mass},\ref{Burgers_T} \& \ref{Burgers_n}), it is obvious that the solution of $\alpha$ and $\beta$ is:
\begin{eqnarray}
&&\left[ {\begin{array}{*{20}{c}}
   \beta  & \alpha   \\
\end{array}} \right] = {\left[ {\begin{array}{*{20}{c}}
   {A - \frac{2}{5}B}  \\
   B  \\
   C^T  \\
   D^T  \\
\end{array}} \right]^{ - 1}}\left[ {\begin{array}{*{20}{c}}
   E & O  \\
   O & {\frac{5}{2n}[n]}  \\
   O & 0  \\
   O & 0  \\
\end{array}} \right],
\end{eqnarray}%
where $O$ is matrix or vector with all elements being zero and $E$ is the identity matrix.

By using $\alpha$, $\beta$, and assuming full ionization, the diffusion velocity can be represented as a linear combination of the gradients of pressure, temperature and abundance as follow:
\begin{eqnarray}
&&{w_j} = {s_j} = P({\alpha _j} - \mu {c_j})\frac{{d\ln T}}{{dr}} + \mu P{c_j}\frac{{d\ln P}}{{dr}} \nonumber \\
&&-\mu P\sum\limits_{i = 1}^{I - 1} {\frac{{{1}}}{{{A_i}}}[{c_j}\mu (1 + {Z_i}) - ({\beta _{ji}} + {\beta _{jI}}{Z_i})]} \frac{{d{X_i^{(0)}}}}{{dr}}, \label{Dif_flux}
\end{eqnarray}%
with
\begin{eqnarray}
&&{c_j} = \sum\limits_{l = 1}^{I - 1} {({\beta _{jl}} + {\beta _{jI}}{Z_l})} \frac{{{X_l^{(0)}}}}{{{A_l^{(0)}}}}
\end{eqnarray}%
and the average molecular weight
\begin{eqnarray}
&&\mu  = [{{\sum\limits_{i = 1}^{I - 1} {\frac{{{X_i^{(0)}}(1 + {Z_i})}}{{{A_i^{(0)}}}}} }}]^{-1},
\end{eqnarray}%
where $X_i^{(0)}$ is the mass abundance of the $i$-th \emph{element} and ${A_i^{(0)}}$ is the relative atomic mass of the $i$-th \emph{element}.

The diffusion equation for each \emph{species} is
\begin{eqnarray}
&&\frac{{\partial {X_j}}}{{\partial t}} = \frac{\partial }{{\partial m}}(4\pi \rho {r^2}{w_j}{X_j}),1 \leq j \leq I, \label{dif_eq_particle}
\end{eqnarray}%
where ${X_j}$ is the mass abundance of the $j$-th \emph{species} ($j<I$ for ions and $j=I$ for electrons). For the $j$-th element ($j<I$), in the case of full ionization, the relation between its element mass abundance $X_j^{(0)}$ and its ion mass abundance $X_j$ is:
\begin{eqnarray}
&&X_j^{(0)} = \frac{{{A_j^{(0)}}}}{{A_j}}{X_j} = \frac{{{A_j^{(0)}}}}{{{A_j^{(0)}-Z_j A_I}}}{X_j}.
\end{eqnarray}%
Therefore, the diffusion equation for each \emph{element} as follow can be obtained by multiplying the constant ${{{A_j^{(0)}}}}/{{A_j}}$ on Eq.(\ref{dif_eq_particle}):
\begin{eqnarray}
&&\frac{{\partial {X_j}^{(0)}}}{{\partial t}} = \frac{\partial }{{\partial m}}(4\pi \rho {r^2}{w_j}{X_j}^{(0)}),1 \leq j \leq I-1. \label{dif_eq_X}
\end{eqnarray}%

\subsection{in solar model}

I have tested the resistance coefficients derived in this paper in Burgers equation in the solar model. The YNEV stellar evolutionary code \citep{zqs15} is used to calculate the stellar models.
The thermodynamic functions are interpolated from the OPAL equation-of-state (EOS) tables EOS2005 \citep{EOS2005}.
The opacities are interpolated from the OPAL tables \citep{OPAC} in high temperature range and \citeauthor{OPLT}'s (\citeyear{OPLT}) tables in low temperature range.
Both the EOS and opacities are interpolated by using bi-cubic polynomials on density and temperature. The nuclear reactions rates are based on \citet{Nucl99} and \citet{CF88} and enhanced by weak electron screening \citep{sal54}. The adopted metal composition is the solar composition by \citet[AGSS09]{AGSS09}. The mixing-length theory parameter $\alpha_{\rm{MLT}}$, initial helium abundance $Y_{ini}$ and initial metallicity $Z_{ini}$ are iteratively adjusted to calibrate the luminosity, radius and the surface abundance ratio $(Z/X)_s$ of the solar models at present age $4.57{\rm{Gyr}}$ to achieve $L=3.8418 \times 10^{33} \rm{erg/s}$ \citep{fro98,bah05}, $R=6.9566 \times 10^{10} \rm{cm}$ \citep{bc98,hsk08} and $(Z/X)_s=0.0181$ \citep{AGSS09}. The K-S $T-\tau$ relation \citep{ks66} is adopted in the atmosphere in solar models.

In the calculations of solar models, two approaches of modeling micro diffusion are compared with each other. One is to use the screening resistance coefficients derived in this paper and another is same to \citet[][CGK89]{cgk89} and \citet[][TBL94]{tbl94} who have used the following resistance coefficients:
\begin{eqnarray}
&&{K_{ij}} = \frac{{1.6249\ln (1 + 0.18769{\Lambda _{ij}}^{1.2})}}{{{\rm{ln}}(1+\Lambda _{ij}^2)}}{K_{ij}^{(0)}},
\end{eqnarray}%
$z_{ij}=0.6$, $z_{ij}'=1.3$ and $z_{ij}''=2$. In CGK89 and TBL94, the adopted resistance coefficient ${K_{ij}}$ is \citeauthor{im85}'s (\citeyear{im85}) analytical fitting of screening resistance coefficient (only in the case of repulsive potential) of \citeauthor{fm79}'s (\citeyear{fm79}) results, but the values of $z_{ij}$, $z_{ij}'$ and $z_{ij}''$ are based on the truncated pure Coulomb potential case. By using resistance coefficients and the solution of Burgers equation (Eq.\ref{Dif_flux}), and taking into account the convective mixing by add a convective diffusion flux term $D_{\rm{conv}}dX_j^{(0)}/dr$ on the r.h.s. of Eq.(\ref{Dif_flux}) in convection zone with $D_{\rm{conv}}=10^{10}{\rm{cm}}^2/{\rm{s}}$ which is large enough to ensure fully mixed convection zone, the equation for elements diffusion (Eq.\ref{dif_eq_X}) with zero flux boundary conditions ($w_j=0$) at the center and the surface of the stellar model is numerically solved.

\begin{figure}
\centering
\includegraphics[scale=0.6]{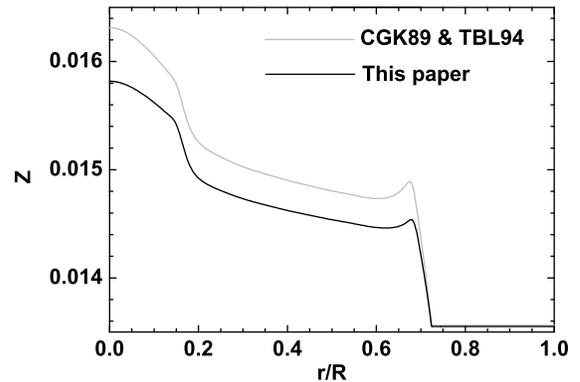}
\caption{ The profile of metallicity of two solar models with different resistance coefficients. \label{solar_z}}
\end{figure}

\begin{figure}
\centering
\includegraphics[scale=0.6]{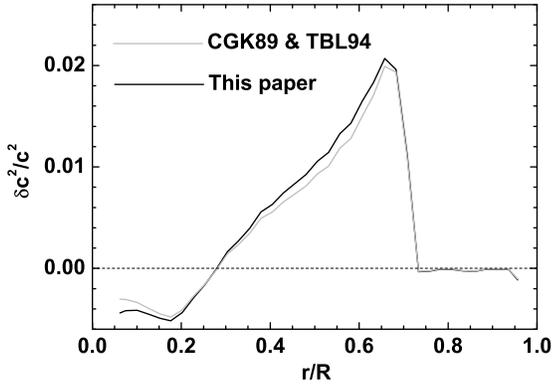}
\caption{ The differences of solar sound speed between models with different resistance coefficients and helioseismic suggested values based on \citet{ba09}. \label{solar_vs}}
\end{figure}

Table \ref{solarinf} shows the information of calibrated solar models with different resistance coefficients. It is found that there are small differences between the results. The values of $\Delta Y_s$, the variation of the surface helium abundance, show that the effect of the diffusion in the solar model with screening resistance coefficients is lower the solar model with resistance coefficients adopted in CGK89 and TBL94. This can be also found in Fig.\ref{solar_z}, which shows that the model with CGK89 and TBL94 resistance coefficients has higher metallicity in the interior of the solar model. This result is same to \citet{pm91}, who compared solar models with resistance coefficients adopted in CGK89, PPFM86 and \citet{n77}. Lower effect of diffusion leads to a higher $R_{bc}$ and a higher $Y_s$ as shown in Table \ref{solarinf}, and the worse $R_{bc}$ leads to a worse sound speed profile as shown in Fig.\ref{solar_vs}.

\begin{table}
\caption{ Information of calibrated solar models with different resistance coefficients. \label{solarinf}}
\centering
\begin{tabular}{lccc}
    \\ \hline
    $  $ & helio.  &  This paper  &  CGK89 \& TBL94
    \\ \hline
    $\alpha_{\rm{MLT}}$ & $-$ & $2.301$ & $2.313$  \\
    $Y_{ini}$ & $-$ & $0.2659$ & $0.2677$  \\
    $Z_{ini}$ & $-$ & $0.01479$ & $0.01511$  \\
    $Y_s$ & $0.2485(35)^{\rm{a}}$ & $0.2376$ & $0.2371$  \\
    $Z_s$ & $-$ & $0.01356$ & $0.01355$  \\
    $R_{bc}$ & $0.7135(5)^{\rm{b,c,d,e}}$ & $0.7245$ & $0.7237$  \\
    $ \Delta Y_s $ & $-$ & $0.0283$ & $0.0306$  \\
\hline
\end{tabular}
\tablecomments{'helio.' means the values suggested by helioseismic studies. $ \Delta Y_s = Y_{ini} - Y_s$ is the variation of helium abundance at the surface. References: a - \citet{ba95,ba04}, b - \citet{chr91}, c - \citet{gc93}, d - \citet{ba97}, e - \citet{ba98}. }
\end{table}

The discrepancies between the solar model and helioseismic studies on $R_{bc}$, $Y_s$, sound speed profile and density profile are called the 'solar abundance problem', since solar models with old high-metallicity compositions \citep[i.e.,][]{GN93,GS98} are consistent with helioseismic observations but the solar model with new low-metallicity compositions AGSS09 is not. A proposal to solve the problem is to enhance the micro diffusion in solar interior \citep[e.g.,][]{mon04,gwc05,yb07,y16}, which increases metallicity in the interior of solar models thus increases the depth of the base of the convection zone and leads to better agreement on the sound speed profile. They found that multiplying a factor $\sim 2$ on TBL94 diffusion velocities can obtain a correct $R_{bc}$. However, such a large factor is physically unacceptable, and the diffusion velocity should actually \emph{decrease} if the screening is taken into account as shown in this paper. On another hand, although enhancing the micro diffusion could result in a correct $R_{bc}$, it makes $Y_s$ worse. A recent study on the solar convective envelope \citep{zqs14} has shown that, with nominal input physics, $R_{bc}$, $Y_s$ and the density profile in the convective envelope cannot be consistent with helioseismic observations in a same time. Because the structure of the solar convective envelope can be determined without stellar evolution calculations, only using modifications in the solar radiative core or on in the chemical evolution equations can not solve the 'solar abundance problem' \citep{zqs14}. Therefore only to modify micro diffusion velocities can not solve the problem.

\section{Summary} \label{summary}

I have computed the collision integrals which determines the resistance coefficients in Burgers equation for diffusion in screening case. The accuracy of the results are high ($< 10^{-12}$ by using Romberg¡¯s method or $< 10^{-10}$ by using fifth order Hermite interpolation from the data in Tables \ref{Fp11data}-\ref{Fa22data} in Appendix \ref{appendix2}). The data in the tables cover $-12 \leq \psi \leq 5$ which should be enough for applications. For very weak screening case $\psi \geq 4.5$, analytical fittings, i.e., Eqs.(\ref{F1_fit}-\ref{F4_fit}), have been obtained with an accuracy of $<10^{-11}$.

The comparison of the resulting resistance coefficients among different works and this paper shows that, in the repulsive potential case, \citeauthor{ppfm86}'s (\citeyear{ppfm86}) results are consistent with this paper, however, in the intermediate and strong screening attractive potential case, the differences are significant and their results seem numerical unstable. For the weak screening limit, their results show small differences comparing with the truncated pure Coulomb potential results, but the results in this paper are exactly reproduce the truncated pure Coulomb potential results.

The resulting resistance coefficients and Burgers equation have been tested in the solar model. It is found that the screening resistance coefficients leads to lower effect of diffusion velocity comparing with the resistance coefficients adopted in TBL94. This is contrary to the expected enhanced diffusion in the solar models.

\begin{acknowledgments}

\textsl{Acknowledgments.}
Many thanks to the anonymous referee for careful reading of the manuscript and comments which improved the original version. 
Fruitful discussions with Y. H. Chen are appreciated. This work is co-sponsored by the National Natural Science Foundation of China through grant No.11303087 and the Chinese Academy of Sciences (¡°Light of West China¡± Program and Youth Innovation Promotion Association).

\end{acknowledgments}

\begin{appendix}

\section{A. Analysis of the properties of a transcendental equation} \label{appendix1}

In the integral of $\Phi$, the independent variable has been changed from $B$ to $R_0$. It is necessary to investigate the interval of $R_0$.

Defining a transcendental function
\begin{eqnarray}
&&H(a,R) = {R^2} - \frac{4}{a}R\exp ( - R),
\end{eqnarray}%
the closest approach for given impact parameter $B$ is determined by $H(a,R_0)=B^2$.

For the case of $a>0$ (corresponding to repulsive potential, $Z_iZ_j>0$), let $H_1(a,R,B)=[H(a,R)-B^2]/R$. There are $H_1(a,0^+,B) < 0$, $H_1(a,+\infty,B) \rightarrow +\infty$ and $ \partial H_1 / \partial R >0 $. Therefore there is only one root $R_0>0$ satisfying $H_1(a,R_0,B)=0$ for an arbitrary $B^2\geq0$. Accordingly,
\begin{eqnarray}
&&\frac{{\partial R_0}}{{\partial B}} =  - \frac{{{{\partial {H_1(a,R_0,B)}}}/{{\partial B}}}}{{{{\partial {H_1(a,R_0,B)}}}/{{\partial R_0}}}}  \\ \nonumber
&&= \frac{{2B}}{{R_0[1 + 4\exp ( - R_0)/a + B^2/R_0^2]}} \ge 0,
\end{eqnarray}%
which also means ${{\partial R}}/{{\partial B}}$ always exists when $R>0$.
Therefore $R_0$ is a continuous and monotonically increasing function of $B$.

For the case of $a<0$ (corresponding to attractive potential, $Z_iZ_j<0$), the situation is more complicated. Analysis of the properties of the low-order derivatives of $H(a,R)$ could help to understand its behavior. The first, second and third order derivatives of $H(a,R)$ with respect to $R$ and their basic properties are as follows:
\begin{eqnarray}
&&H' = 2R + \frac{4}{a}(R - 1)\exp ( - R), \label{H'0} \\
&&H'' = 2 + \frac{4}{a}(2 - R)\exp ( - R), \label{H''0} \\
&&H''(0) = 2(1 + \frac{4}{a}), \label{H''1} \\
&&H''(R \geq 2) \geq 2, \label{H''2} \\
&&H''' = \frac{4}{a}(R - 3)\exp ( - R), \label{H'''0} \\
&&H'''(R < 3) > 0. \label{H'''3}
\end{eqnarray}%

\begin{figure*}
\centering
\includegraphics[scale=0.6]{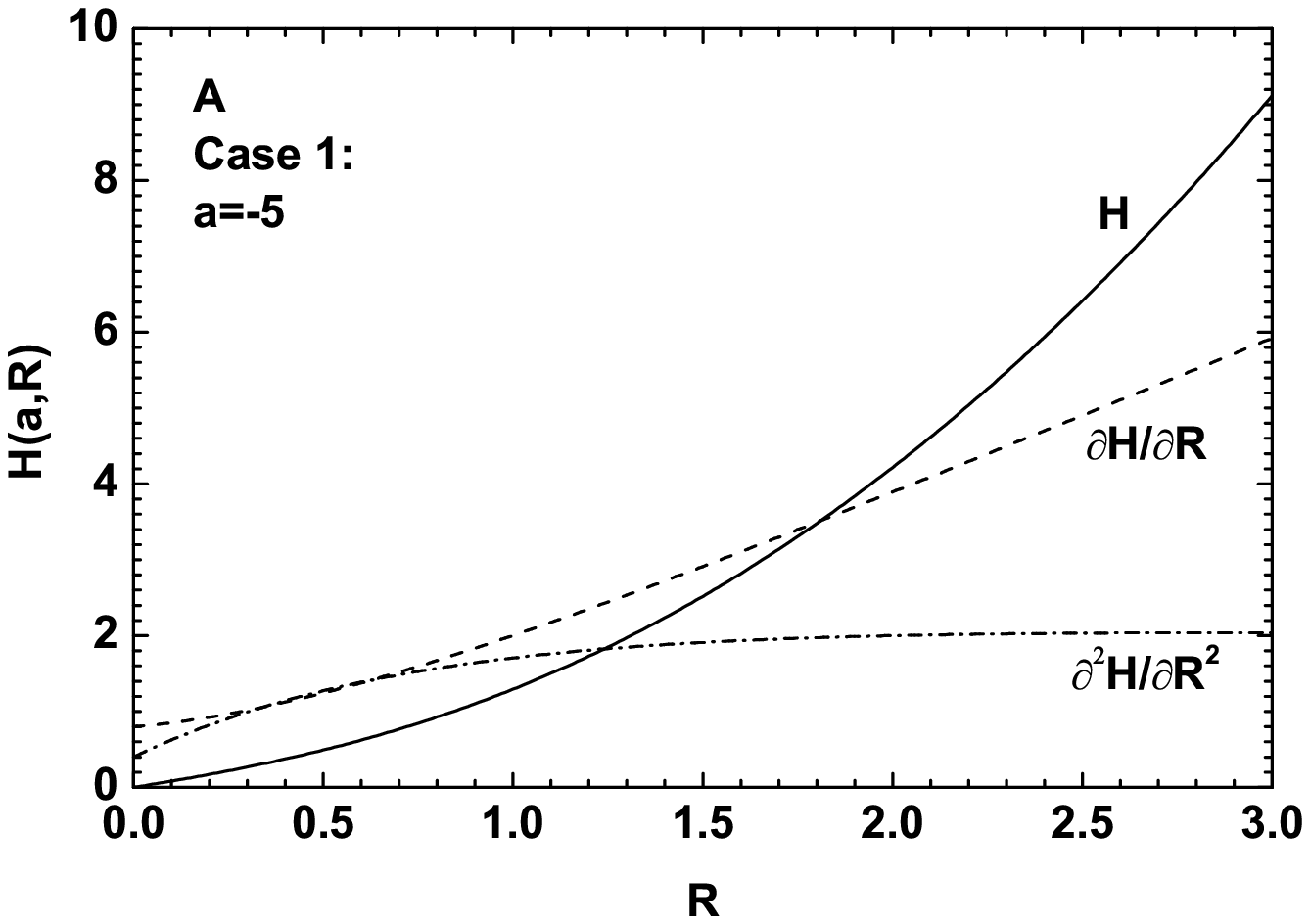}
\includegraphics[scale=0.6]{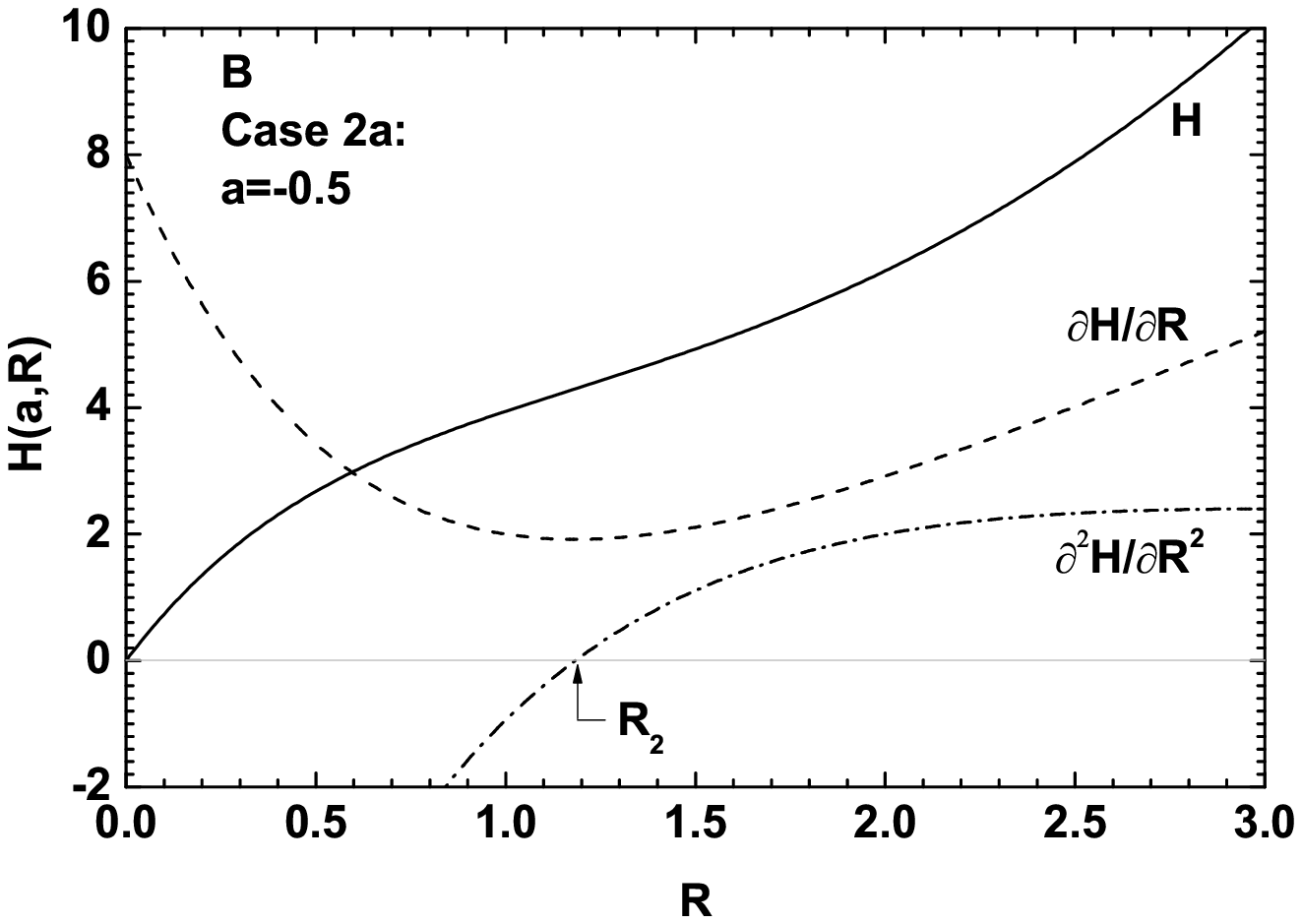}
\includegraphics[scale=0.6]{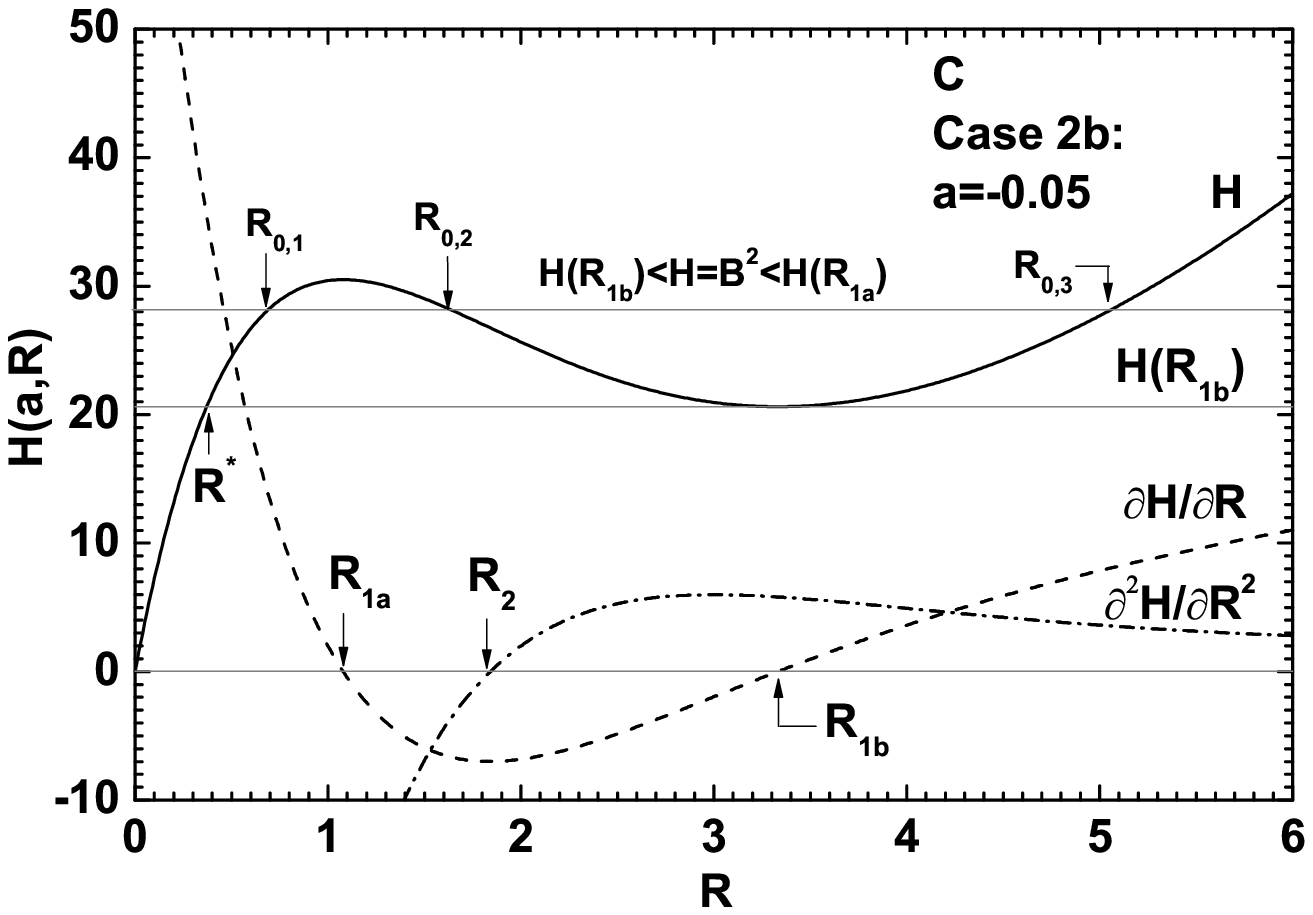}
\includegraphics[scale=0.6]{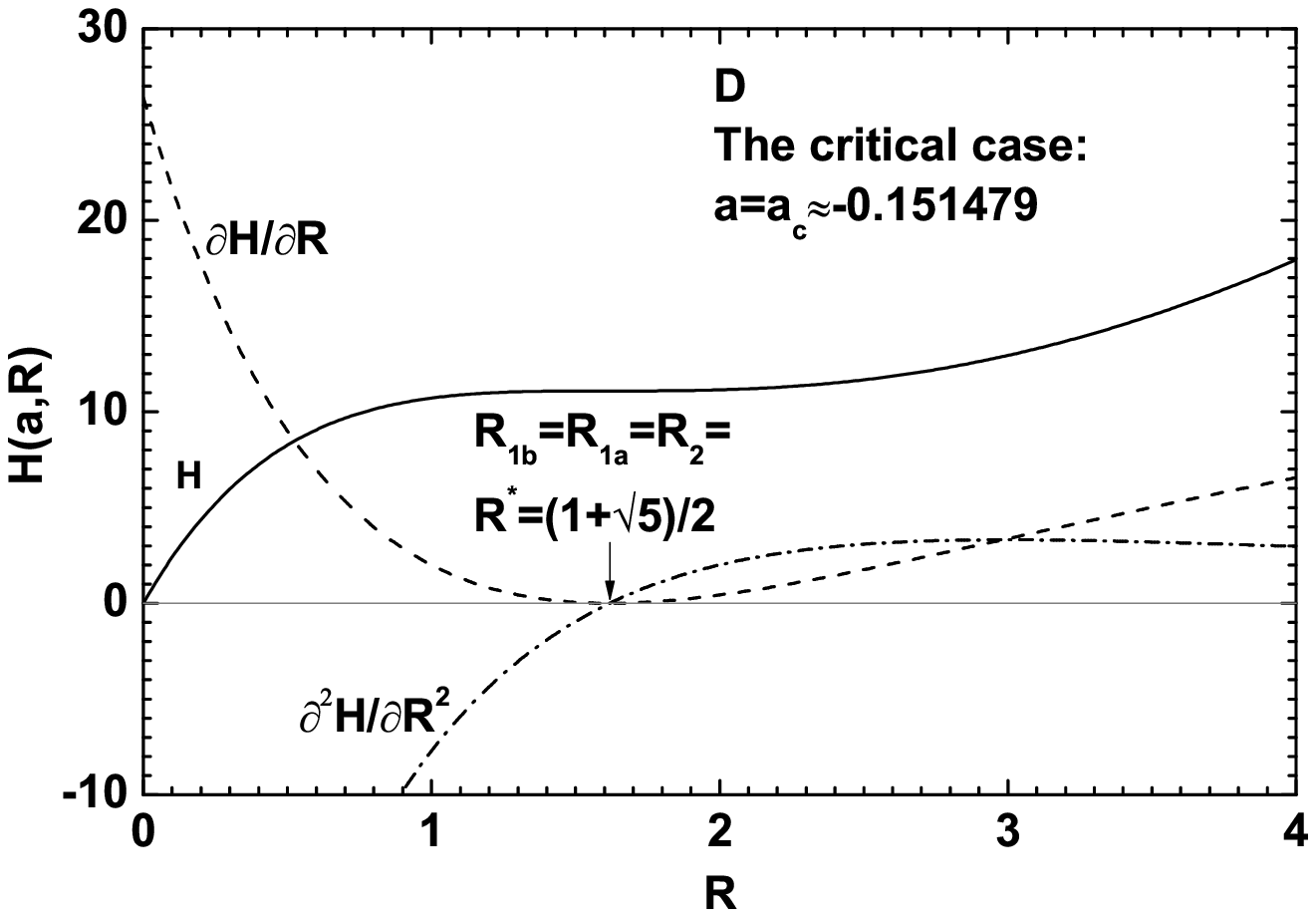}
\caption{$H(a,R)$, $\partial H/\partial R$ and $\partial ^2 H/\partial R^2$ for different $a<0$ in four cases. \label{AppendixCase}}
\end{figure*}

The behaviors of $H$ can be analyzed as follows:

According to Eq.(\ref{H'''3}), $H''$ is monotonically increasing in $[0,3]$.

Case 1 (see Fig.\ref{AppendixCase}A): if $4/a\geq-1$, $H''(0<R<2)>H''(0)\geq0$. By taking into account Eq.(\ref{H''2}), there is $H''>0$. Therefore $H' \geq H'(0)>0$ and $H$ is monotonically increasing. There is only one root $R_0 \geq 0$ satisfy $H(a,R_0)=B^2$ for arbitrary $B \geq 0$.

Case 2: if $4/a<-1$, there is only one root for $H''=0$ and the root is in $(0,2)$, defining the root as $R_2$. Since $H'''(R_2)>0$, it is obvious that $H'(R_2)$ is the minimum of $H'$.

Case 2a (see Fig.\ref{AppendixCase}B): if $H'(R_2)\geq0$, $H'\geq H'(R_2)\geq0$. $H$ is monotonically increasing, which is similar to Case 1.

Case 2b (see Fig.\ref{AppendixCase}C): if $H'(R_2)<0$, because $H'(0)>0$ and $H'(+\infty)\rightarrow +\infty$, there are two roots for $H'(R)=0$. There can not be three roots or more because $H''=0$ has only one root. The two roots of $H'(R)=0$ are defined as $R_{1a}$ and $R_{1b}$, which satisfy $R_{1a}<R_2<R_{1b}$. $H'$ satisfies $H'(R<R_{1a})>0$, $H'(R_{1a}<R<R_{1b})<0$ and $H'(R>R_{1b})>0$. Because $H''(R_2)=0$ is the only root of $H''$ and $H''$ is monotonically increasing in $[0,3]$, there are $H''(R_{1a})<0$ and $H''(R_{1b})>0$, thus $H(R_{1a})$ is a local maximum of $H$ and $H(R_{1b})$ is a local minimum. Since $H(0)=0$ and $H'(R<R_{1a})>0$, there is $H(R_{1a})>H(R_{1b})>0$, therefore there is only one $R^*$ satisfying $0<R^*<R_{1a}$ and $H(R^*)=H(R_{1b})$.

In case 2b, for continuous changing $B^2$ with $B^2<H(R^*)$, the solution $R_0$ for $B^2=H(R_0)$ continuously changes from $R_0=0$ to $R_0=R^*$. For continuous changing $B^2$ with $B^2>H(R^*)=H(R_{1b})$, the solution $R_0$ continuously changes from $R_0=R_{1b}$ to $R_0\rightarrow+\infty$. In the latter case, there could be three roots for $B^2=H(R_0)$ when $H(R_{1b})<B^2<H(R_{1a})$: $R^*<R_{0,1}<R_{1a}<R_{0,2}<R_{1b}<R_{0,3}$. However, the physical acceptable root of the closet approach $R_0$ is the maximum root $(R_0=R_{0,3})$ which satisfy $R_0>R_{1b}$. The explanation is as follow. See Fig.\ref{AppendixCase}C for example, for an approaching particle with impact parameter $B$ satisfying $H(R_{1b})<B^2<H(R_{1a})$, the particle existing in the regions in which $H>B^2$ (e.g., $R_{0,1}<R<R_{0,2}$ and $R>R_{03}$) is allowed in the view of energy. On the other hand, in the view of classical kinematics, the particle can not pass the barrier zone ($R_{0,2}<R<R_{0,3}$) into the allowed existing region ($R_{0,1}<R<R_{0,2}$). Therefore the closest approach for the particle is $R_{0,3}$. As a result, in case 2b, for an arbitrary $B$, the corresponding closest approach $R_0$ can never be in $(R^*,R_{1b})$.

Accordingly, the critical condition between case 2a and case 2b is $H'(R_2)=0$ (see Fig.\ref{AppendixCase}D). The critical value of $a$ and $R_2$ can be worked out by using $H''(R_2)=0$ and $H'(R_2)=0$:
\begin{eqnarray}
&&2{R_{2,c}} + \frac{4}{a_c}({R_{2,c}} - 1)\exp ( - {R_{2,c}}) = 0
\end{eqnarray}%
and
\begin{eqnarray}
&&2 + \frac{4}{a_c}(2 - {R_{2,c}})\exp ( - {R_{2,c}}) = 0,
\end{eqnarray}%
where $a_c$ is the critical values of $a$ for case 2b and $R_{2,c}$ is the corresponding $R_2$ for $a=a_c$.
The solution is $R_{2,c}=(1+\sqrt{5})/2$ and $a_c=- (3 - \sqrt 5 )\exp [ - (1 + \sqrt 5 )/2] \approx -0.151479$.

\section{B. Tables of collision integrals and their derivatives} \label{appendix2}

\begin{longtable}{lccc}
\caption{ ${\rm{ln}}F^{(11)}$ and its derivative for repulsive potential ($\Lambda>0$). \label{Fp11data}}
    \\ \hline
    $\Lambda > 0$ & &
    \\ \hline
    $\psi$ & ${\rm{ln}}F^{(11)}$ & $ d {\rm{ln}}F^{(11)} / d \psi $ & $ d^2 {\rm{ln}}F^{(11)} / d \psi^2 $
    \\ \hline
    \endhead
    $  -7.0$ & $  -6.0501726540$ & $   0.7692860908$ & $  -0.0192141113$  \\
    $  -6.9$ & $  -5.9733405036$ & $   0.7673530333$ & $  -0.0194471810$  \\
    $  -6.8$ & $  -5.8967028256$ & $   0.7653966305$ & $  -0.0196809717$  \\
    $  -6.7$ & $  -5.8202619576$ & $   0.7634168248$ & $  -0.0199151864$  \\
    $  -6.6$ & $  -5.7440202416$ & $   0.7614135901$ & $  -0.0201494932$  \\
    $  -6.5$ & $  -5.6679800203$ & $   0.7593869354$ & $  -0.0203835216$  \\
    $  -6.4$ & $  -5.5921436336$ & $   0.7573369088$ & $  -0.0206168592$  \\
    $  -6.3$ & $  -5.5165134146$ & $   0.7552636018$ & $  -0.0208490471$  \\
\hline
\end{longtable}

\begin{longtable}{lccc}
\caption{ ${\rm{ln}}F^{(12)}$ and its derivative for repulsive potential ($\Lambda>0$). \label{Fp12data}}
    \\ \hline
    $\Lambda > 0$ & &
    \\ \hline
    $\psi$ & ${\rm{ln}}F^{(12)}$ & $ d {\rm{ln}}F^{(12)} / d \psi $ & $ d^2 {\rm{ln}}F^{(12)} / d \psi^2 $
    \\ \hline
    \endhead
    $  -7.0$ & $  -5.1188471542$ & $   0.7538748345$ & $  -0.0215462370$  \\
    $  -6.9$ & $  -5.0435678628$ & $   0.7517063774$ & $  -0.0218231937$  \\
    $  -6.8$ & $  -4.9685068046$ & $   0.7495101427$ & $  -0.0221017429$  \\
    $  -6.7$ & $  -4.8936667650$ & $   0.7472859849$ & $  -0.0223816068$  \\
    $  -6.6$ & $  -4.8190505423$ & $   0.7450337879$ & $  -0.0226624738$  \\
    $  -6.5$ & $  -4.7446609449$ & $   0.7427534683$ & $  -0.0229439952$  \\
    $  -6.4$ & $  -4.6705007877$ & $   0.7404449799$ & $  -0.0232257816$  \\
    $  -6.3$ & $  -4.5965728881$ & $   0.7381083176$ & $  -0.0235073993$  \\
\hline
\end{longtable}

\begin{longtable}{lccc}
\caption{ ${\rm{ln}}F^{(13)}$ and its derivative for repulsive potential ($\Lambda>0$). \label{Fp13data}}
    \\ \hline
    $\Lambda > 0$ & &
    \\ \hline
    $\psi$ & ${\rm{ln}}F^{(13)}$ & $ d {\rm{ln}}F^{(13)} / d \psi $ & $ d^2 {\rm{ln}}F^{(13)} / d \psi^2 $
    \\ \hline
    \endhead
    $  -7.0$ & $  -3.8640683807$ & $   0.7413971954$ & $  -0.0234601900$  \\
    $  -6.9$ & $  -3.7900464818$ & $   0.7390355760$ & $  -0.0237725907$  \\
    $  -6.8$ & $  -3.7162623107$ & $   0.7366426030$ & $  -0.0240872180$  \\
    $  -6.7$ & $  -3.6427190134$ & $   0.7342180671$ & $  -0.0244038029$  \\
    $  -6.6$ & $  -3.5694197554$ & $   0.7317617872$ & $  -0.0247220433$  \\
    $  -6.5$ & $  -3.4963677191$ & $   0.7292736145$ & $  -0.0250416006$  \\
    $  -6.4$ & $  -3.4235660995$ & $   0.7267534358$ & $  -0.0253620963$  \\
    $  -6.3$ & $  -3.3510181013$ & $   0.7242011780$ & $  -0.0256831085$  \\
\hline
\end{longtable}

\begin{longtable}{lccc}
\caption{ ${\rm{ln}}F^{(22)}$ and its derivative for repulsive potential ($\Lambda>0$). \label{Fp22data}}
    \\ \hline
    $\Lambda > 0$ & &
    \\ \hline
    $\psi$ & ${\rm{ln}}F^{(22)}$ & $ d {\rm{ln}}F^{(22)} / d \psi $ & $ d^2 {\rm{ln}}F^{(22)} / d \psi^2 $
    \\ \hline
    \endhead
    $  -7.0$ & $  -5.1022738271$ & $   0.7797893985$ & $  -0.0189772192$  \\
    $  -6.9$ & $  -5.0243901918$ & $   0.7778791155$ & $  -0.0192287312$  \\
    $  -6.8$ & $  -4.9466988452$ & $   0.7759435986$ & $  -0.0194818528$  \\
    $  -6.7$ & $  -4.8692023182$ & $   0.7739827002$ & $  -0.0197363146$  \\
    $  -6.6$ & $  -4.7919031552$ & $   0.7719963011$ & $  -0.0199918127$  \\
    $  -6.5$ & $  -4.7148039109$ & $   0.7699843145$ & $  -0.0202480052$  \\
    $  -6.4$ & $  -4.6379071470$ & $   0.7679466897$ & $  -0.0205045079$  \\
    $  -6.3$ & $  -4.5612154280$ & $   0.7658834168$ & $  -0.0207608906$  \\
\hline
\end{longtable}

\begin{longtable}{lccc}
\caption{ ${\rm{ln}}F^{(11)}$ and its derivative for attractive potential ($\Lambda<0$). \label{Fa11data}}
    \\ \hline
    $\Lambda < 0$ & &
    \\ \hline
    $\psi$ & ${\rm{ln}}F^{(11)}$ & $ d {\rm{ln}}F^{(11)} / d \psi $ & $ d^2 {\rm{ln}}F^{(11)} / d \psi^2 $
    \\ \hline
    \endhead
    $  -7.0$ & $  -5.5320878941$ & $   0.7997301452$ & $  -0.0178786640$  \\
    $  -6.9$ & $  -5.4522017023$ & $   0.7980221339$ & $  -0.0161703872$  \\
    $  -6.8$ & $  -5.3724766176$ & $   0.7965199232$ & $  -0.0137471302$  \\
    $  -6.7$ & $  -5.2928883397$ & $   0.7952992773$ & $  -0.0105277962$  \\
    $  -6.6$ & $  -5.2134046321$ & $   0.7944426445$ & $  -0.0064613444$  \\
    $  -6.5$ & $  -5.1339848199$ & $   0.7940357066$ & $  -0.0015354812$  \\
    $  -6.4$ & $  -5.0545796721$ & $   0.7941631354$ & $   0.0042163044$  \\
    $  -6.3$ & $  -4.9751317408$ & $   0.7949037589$ & $   0.0107100792$  \\
\hline
\end{longtable}

\begin{longtable}{lccc}
\caption{ ${\rm{ln}}F^{(12)}$ and its derivative for attractive potential ($\Lambda<0$). \label{Fa12data}}
    \\ \hline
    $\Lambda < 0$ & &
    \\ \hline
    $\psi$ & ${\rm{ln}}F^{(12)}$ & $ d {\rm{ln}}F^{(12)} / d \psi $ & $ d^2 {\rm{ln}}F^{(12)} / d \psi^2 $
    \\ \hline
    \endhead
    $  -7.0$ & $  -4.5770645479$ & $   0.7856964332$ & $  -0.0077012431$  \\
    $  -6.9$ & $  -4.4985225877$ & $   0.7852571585$ & $  -0.0008383846$  \\
    $  -6.8$ & $  -4.4199878009$ & $   0.7855771123$ & $   0.0074717442$  \\
    $  -6.7$ & $  -4.3413771879$ & $   0.7867957983$ & $   0.0171069532$  \\
    $  -6.6$ & $  -4.2625945889$ & $   0.7890351051$ & $   0.0278366180$  \\
    $  -6.5$ & $  -4.1835329890$ & $   0.7923884533$ & $   0.0393229663$  \\
    $  -6.4$ & $  -4.1040778876$ & $   0.7969105262$ & $   0.0511315539$  \\
    $  -6.3$ & $  -4.0241116262$ & $   0.8026085180$ & $   0.0627511194$  \\
\hline
\end{longtable}

\begin{longtable}{lccc}
\caption{ ${\rm{ln}}F^{(13)}$ and its derivative for attractive potential ($\Lambda<0$). \label{Fa13data}}
    \\ \hline
    $\Lambda < 0$ & &
    \\ \hline
    $\psi$ & ${\rm{ln}}F^{(13)}$ & $ d {\rm{ln}}F^{(13)} / d \psi $ & $ d^2 {\rm{ln}}F^{(13)} / d \psi^2 $
    \\ \hline
    \endhead
    $  -7.0$ & $  -3.3043094402$ & $   0.7811842652$ & $   0.0263704421$  \\
    $  -6.9$ & $  -3.2260345265$ & $   0.7845661347$ & $   0.0414881580$  \\
    $  -6.8$ & $  -3.1473438908$ & $   0.7895154209$ & $   0.0576060363$  \\
    $  -6.7$ & $  -3.0680769832$ & $   0.7960957359$ & $   0.0739716267$  \\
    $  -6.6$ & $  -2.9880708782$ & $   0.8042889549$ & $   0.0897114159$  \\
    $  -6.5$ & $  -2.9071689593$ & $   0.8139860162$ & $   0.1038915204$  \\
    $  -6.4$ & $  -2.8252301964$ & $   0.8249844622$ & $   0.1155907598$  \\
    $  -6.3$ & $  -2.7421382924$ & $   0.8369936033$ & $   0.1239785828$  \\
\hline
\end{longtable}

\begin{longtable}{lccc}
\caption{ ${\rm{ln}}F^{(22)}$ and its derivative for attractive potential ($\Lambda<0$). \label{Fa22data}}
    \\ \hline
    $\Lambda < 0$ & &
    \\ \hline
    $\psi$ & ${\rm{ln}}F^{(22)}$ & $ d {\rm{ln}}F^{(22)} / d \psi $ & $ d^2 {\rm{ln}}F^{(22)} / d \psi^2 $
    \\ \hline
    \endhead
    $  -7.0$ & $  -4.8821674499$ & $   0.8127115778$ & $  -0.0056399620$  \\
    $  -6.9$ & $  -4.8009206265$ & $   0.8122645965$ & $  -0.0032600530$  \\
    $  -6.8$ & $  -4.7197062732$ & $   0.8120647469$ & $  -0.0007268170$  \\
    $  -6.7$ & $  -4.6384992303$ & $   0.8121175184$ & $   0.0017532055$  \\
    $  -6.6$ & $  -4.5572749137$ & $   0.8124050282$ & $   0.0039212121$  \\
    $  -6.5$ & $  -4.4760118901$ & $   0.8128815522$ & $   0.0054825127$  \\
    $  -6.4$ & $  -4.3946948039$ & $   0.8134710380$ & $   0.0061295089$  \\
    $  -6.3$ & $  -4.3133174305$ & $   0.8140671585$ & $   0.0055690923$  \\
\hline
\end{longtable}

\end{appendix}

\end{document}